\documentclass[12pt,preprint]{aastex}

\def\equ#1{eq.~(\ref{eq:#1})}
\def\se#1{\S\ref{sec:#1}}
\def\Fig#1{Fig.~\ref{fig:#1}}

\def\\{\hfill\break}

\def\etal{{\it et al.\ }}

\def\la{\langle}
\def\ra{\rangle}
\def\be{\begin{equation}}
\def\ee{\end{equation}}
\def\prop{\propto}
\def\ifm#1{\relax\ifmmode#1\else$\mathsurround=0pt #1$\fi}
\def\kms{\ifmmode\,{\rm km}\,{\rm s}^{-1}\else km$\,$s$^{-1}$\fi}

\def\dd{{\rm d}}

\def\msun{M_{\odot}}

\def\ltsima{$\; \buildrel < \over \sim \;$}
\def\lsim{\lower.5ex\hbox{\ltsima}}
\def\gtsima{$\; \buildrel > \over \sim \;$}
\def\gsim{\lower.5ex\hbox{\gtsima}}

\def\Mv{M_{\rm v}}
\def\mv{m_{\rm v}}
\def\Rv{R_{\rm v}}
\def\Dv{\Delta_{\rm v}}

\def\ai{\alpha_{\rm in}}
\def\ain{\alpha_{\rm in}}
\def\ao{\alpha_{\rm out}}
\def\aout{\alpha_{\rm out}}
\def\a0{\alpha_0}
\def\aas{\alpha_{\rm as}}
\def\ellt{\ell_{\rm t}}
\def\rt{r_{\rm t}}

\def\rs{r_{\rm s}}

\def\ellc{\ell_{\rm c}}

\def\rc{r_{\rm s}}
\def\ellc{\ell_{\rm s}}
\def\rhoc{\rho_{\rm s}}
\def\sigc{\sigma_{\rm s}}

\def\mf{m_{\rm f}}

\def\mnu{\nu}

\def\bFt{{\bf F}_{\rm t}}
\def\br{{\bf r}}

\def\rapo{r_{\rm a}}
\def\rperi{r_{\rm p}}
\def\vcirc{V_{\rm circ}}
\def\tini{t_{\rm i}}
\def\tfric{t_{\rm fric}}
\def\ffric{f_{\rm fric}}
\def\mmin{m_{\rm min}}

\def\pmb#1{\setbox0=\hbox{#1}%
\kern-.025em\copy0\kern-\wd0
\kern.05em\copy0\kern-\wd0
\kern-.025em\raise.0433em\box0}
\def\bell{\pmb{$\ell$}}

\def\rar{\rightarrow}

\begin{document}


\title{DARK-HALO CUSP: ASYMPTOTIC CONVERGENCE}

\author{Avishai Dekel, Itai Arad, Jonathan Devor \& Yuval Birnboim}
\affil{Racah Institute of Physics, The Hebrew University, Jerusalem
91904, Israel}

\begin{abstract}
We propose a model for how the buildup of dark halos by merging satellites 
produces a characteristic inner cusp, of a density profile $\rho\prop r^{-\ai}$
with $\ai\rar\aas\gsim 1$, as seen in cosmological N-body simulations 
of hierarchical clustering scenarios. 
Dekel, Devor \& Hetzroni (2003) argue that a flat core of $\ai<1$ exerts tidal 
compression which prevents local deposit of satellite material; the satellite 
sinks intact into the halo center thus causing a rapid steepening to $\ai>1$.  
Using merger N-body simulations, we learn that this cusp is stable
under a sequence of mergers, and derive a practical tidal mass-transfer 
recipe in regions where the local slope of the halo profile is $\alpha>1$. 
According to this recipe, the ratio of mean densities of halo and initial 
satellite within the tidal radius equals a given function $\psi(\alpha)$, which
is significantly smaller than unity (compared to being $\sim 1$ according to 
crude resonance criteria) and is a {\it decreasing} function of $\alpha$. This 
decrease makes the tidal mass transfer relatively more efficient at larger 
$\alpha$, which means steepening when $\alpha$ is small
and flattening when $\alpha$ is large, thus causing converges to a stable 
solution.  Given this mass-transfer recipe, linear perturbation analysis, 
supported by toy simulations, shows that a sequence of cosmological mergers 
with homologous satellites slowly leads to a fixed-point cusp with an 
asymptotic slope $\aas>1$.  The slope depends only weakly on the fluctuation
power spectrum, in agreement with cosmological simulations.  
During a long interim period the profile has an NFW-like shape, with a cusp of 
$1<\ai<\aas$.  Thus, a cusp is enforced if enough compact satellite remnants 
make it intact into the inner halo.  In order to maintain a flat core, 
satellites must be disrupted outside the core, possibly as a result of 
a modest puffing up due to baryonic feedback.
\end{abstract}

\subjectheadings{
cosmology: theory --- 
dark matter --- 
galaxies: formation --- 
galaxies: halos ---
galaxies: interactions ---
galaxies: kinematics and dynamics}

\section{INTRODUCTION}
\label{sec:intro}

A relatively robust universal shape for the density profile of dark-matter 
halos has been seen in cosmological N-body simulations of dissipationless 
hierarchical clustering from Gaussian initial fluctuations. It can be
approximated by the functional form
\be
\rho (r) = {\rhoc \,
           \left({r\over\rc}\right)^{-\ai} \,
	   \left(1+{r\over\rc}\right)^{\ai-\ao}
	   } \, ,
\label{eq:nfw}
\ee
where $\rc$ and $\rhoc$ are characteristic inner radius and density
respectively.\footnote{a useful generalized functional form has been
proposed by Zhao (1996)} This density profile is characterized by 
an inner ``cusp" $\propto r^{-\ai}$ and a continuous steepening through a bend
near $\rc$ towards $r^{-3}$ near the virial radius $\Rv$
(defined by a fixed mean overdensity $\Dv$ above the universal mean,
with $\Dv =180$ to 340, depending on time and the cosmological model).
Navarro, Frenk \& White (1995; 1996; 1997, hereafter NFW) found
\equ{nfw} with $\ai \simeq 1$ and $\ao \simeq 3$
to be a good fit to halos in simulations
over the radius range $(0.01-1)\Rv$,
for a wide range of halo masses and for a range of hierarchical cosmological
scenarios with different power spectra of initial fluctuations.
Cole \& Lacey (1996) came to a similar conclusion for self-similar scenarios
with power-law power spectra, $P_k\propto k^n$ with $n=0,-1,-2$,
in an Einstein-deSitter cosmology.
High-resolution simulations of a few individual halos in a cosmological
environment (Moore \etal 1998; Ghigna \etal 2000; Klypin \etal 2001)
found that the typical asymptotic cusp profile at $r\ll \rc$ is
sometimes somewhat steeper, closer to $\ai \simeq 1.5$.
A careful convergence analysis by Power \etal (2002), who explored the
robustness to numerical errors,
found for the standard $\Lambda$CDM cosmology that $\ai$ reaches a slope
shallower than 1.2 at their innermost resolved point of $r \sim 0.005\Rv$.
Thus, the robust result of the simulations is that dark halos have
inner cusps with a characteristic slope $1 \leq \ai \leq 1.5$.
We seek a basic theoretical understanding of the origin of these cusps.

An even more intriguing puzzle is introduced by observations which indicate
that at least in some cases the actual inner halo density profiles are
close to flat cores, with $\ai \simeq 0$. This has been detected directly
by rotation curves in low surface brightness (LSB) galaxies,
whose centers are dominated by their dark halos
(van den Bosch \etal 2000; de Block \etal 2001; Marchesini \etal 2002).
Cores have also been argued to exist in normal disk galaxies, where
more involved modeling is required
(Salucci \& Burkert 2000; Salucci 2001; Borriello \& Salucci 2001).
The presence of a core 
seems to introduce a severe challenge to the CDM cosmological paradigm.
We mention in the conclusion section several mechanisms that have been proposed
for the origin of such cores.
In particular,
attempts to turn a cusp into a core by direct stellar feedback effects
in the present halos, which looked promising at a first sight
(Navarro, Eke \& Frenk 1996), seem not to work (e.g., Geyer \& Burkert 2001;
Gnedin \& Zhao 2002).

We find it useful to first try to understand in simple basic terms the 
possible 
origin of the universal cusp in the gravitational N-body simulations of 
cold dark matter.  This 
may 
provide us with a tool for addressing the formation and survival of flat
cores by other mechanisms, in particular by baryonic feedback processes within
the hierarchical CDM framework.

Several mechanisms have been studied in the context of dark-halo profiles in
dissipationless simulations.
The outer slope of $r^{-3}$ (and steeper) may possibly be explained in terms
of violent relaxation (e.g., Barnes \& Hernquist 1991;  
Pearce, Thomas \& Couchman 1993 and references therein).
We note in general that any
finite system would tend to have a steep density fall off at large radii
due to diffusion of particles outwards.
Secondary spherical infall is expected to produce a profile closer to
$\rho \propto r^{-2}$, which may 
explain the behavior in the intermediate regions of the halo, but is
too steep to explain the flatter inner cusp
(Lokas \& Hoffman 2000 and references therein).
Thus, none of the above mechanisms seem to provide a natural explanation for
the characteristic cusp of $\alpha \gsim 1$.

The clear impression from the cosmological 
N-body simulations of hierarchical clustering scenarios is
that halos are largely built
up by a sequence of mergers of smaller structures.\footnote{though 
cusps may also emerge in other scenarions where mergers are suppressed, 
see a discussion in \se{conc}.} 
In a typical merger, a bound 
satellite halo spirals into the center of the larger halo due to gravity
and dynamical friction. The satellite
gradually transfers mass into the host halo due to tidal stripping
or by eventually melting into the halo inner region. 
This process is likely to have an important effect in shaping up the 
density profile.
Indeed, Syer \& White (1998), Nusser \& Sheth (1999) and Subramanian, Cen \&
Ostriker (2000, hereafter SCO) argued, using certain
simple models and simulations of a sequence of mergers, that the buildup by
mergers may naturally lead to a stable profile.  
However, they find their predicted profile to be quite sensitive to the power
spectrum of fluctuations and to allow an inner slope of $\alpha <1$, 
in conflict with the robust result of the cosmological simulations.
In fact, when trying to repeat the Syer \& White analysis using 
their simplified modeling of the stripping process but with higher resolution,
we find that in the long run the profile does not really converge to a
stable cusp but rather continues to steepen slowly towards $\ai=3$ (see
\se{conc}).
Either way, it seems that something is not adequate in the simplified
model adopted to describe the mass transfer from the satellite to the halo.

We re-visit the buildup of halo profile by merging
satellites and gain an encouraging new insight.
We add two important new ingredients to the tidal effects.
In another paper (Dekel, Devor \& Hetzroni 2003, hereafter DDH) 
we argue that for a flat mean density profile 
with $\alpha \leq 1$ the tidal effects on typical satellites induce
three-dimensional compression with no local mass deposit, 
which results in a rapid steepening of the inner
profile to $\alpha > 1$.
In this paper, 
we derive a useful prescription for tidal mass transfer at $\alpha > 1$,
and obtain higher deposit efficiency at higher $\alpha$.  
We then show that this tends to flatten steep profiles with large
$\alpha$ and thus slowly leads to an asymptotic fixed point at a certain
$\alpha = \aas \gsim 1$.

In \se{tide} we address the compression at $\alpha < 1$ and the
resultant steepening (DDH).
In \se{psi} we derive a simple mass-transfer prescription in the range
$\alpha > 1$, based on merger N-body simulations and toy-model understanding.
In \se{qual} we explain why this prescription should lead to an
asymptotic cusp.
In \se{linear} we use linear perturbation analysis to compute the
asymptotic slope for satellites of a given mass, and 
in \se{dist} we extend the analysis to a cosmological distribution of masses.
In \se{toy} we demonstrate this process via a semi-analytic 
simulation of a cosmological sequence of mergers.
In \se{conc} we discuss our results.

\section{CORE TO CUSP BY TIDAL COMPRESSION}
\label{sec:tide}

This introductory section describes the robust transition of a core 
($\ai \leq 1$) to a cusp ($\ai > 1$), which is the main theme of another paper,
DDH. We summarize it here for completeness and as a background
to the independent analysis described in the following sections.

A useful quantity in describing
the tidal forces exerted by a halo of mass profile $M(r)$ is the local 
logarithmic slope of its mean density profile $\bar\rho(r) \prop M(r)/r^3$,
\be
\alpha(r) \equiv - {d\ln \bar\rho \over d\ln r} \,,
\ee
such that locally $\bar\rho \propto r^{-\alpha}$.
We assume that $\alpha$ is either constant or monotonically increasing as
a function of $r$, with values in the range $0 \leq \alpha \leq 3$.
The extreme values of $\alpha=0$ and 3 correspond to a constant-density
halo and a point mass respectively.
Note that if the profile inside $r$ is a power law, then the local and
mean density profiles have the same logarithmic slope. In general,
they are related via $\rho(r) =[1-\alpha(r)/3] \bar{\rho}(r)$,
but they do not necessarily have the same slope at a given $r$.
The slope of $\rho(r)$ is equal to or larger than the slope of $\bar\rho(r)$.
The following analysis refers to $\alpha$ as the slope of $\bar\rho(r)$.

We then consider a satellite of mass $\mv \ll \Mv$, moving under the
gravity excreted by the halo, when its center of mass is at position
$\br$ as measured from the halo center. 
The tidal acceleration exerted by the halo mass distribution on a satellite
particle at position vector $\bell$ relative to the satellite center of mass
is obtained by transforming the gravitational attraction exerted by the halo
on the particle into the (non-rotating) rest frame of the accelerated
satellite. In the tidal limit $\ell \ll r$, to first order in $\ell/r$,
using Cartesian coordinates about the satellite center, where
$\bell = (\ell_1,\ell_2,\ell_3)$ and $\bell_1$ lies along $\br$,
we obtain
\be
{\bFt} = {GM(r) \over r^3}\,
\left( [\alpha(r)-1]\,\bell_1\, -\bell_2 -\bell_3 \right) \,.
\label{eq:tide_harmonic}
\ee

The components perpendicular to the line connecting the centers of mass
are always of compression towards the satellite center, 
with an amplitude that does not explicitly depend on $\alpha$.
The maximum radial tidal force outwards
is obtained along the line connecting the centers of mass.
In the limit where the tides are exerted by a point-mass halo, $\alpha=3$,
the pull outwards is maximal. For flatter halo slopes, the tidal stretching
becomes weaker in proportion to $(\alpha-1)$, until it vanishes at $\alpha=1$
and reverses direction into compression for $\alpha<1$.
Thus, while for $\alpha>1$ there is always a tidal component pulling outwards,
for $\alpha < 1$ the tidal forces are of compression everywhere in the 
satellite, and in the limit of a core with $\alpha=0$ the tides induce 
symmetric compression in all directions.

We argue in DDH that this critical transition at $\alpha \sim 1$ is
the main source for the origin of a cusp steeper than $r^{-1}$.
The idea is that if the local tidal mass transfer from the satellite
to the halo stops when the satellite's orbit has decayed into
a core region where $\alpha(r) \leq 1$, the satellite would
continue to sink in due to dynamical friction without further mass loss 
until it settles in the halo center. 
This would inevitably cause a general steepening of the core profile towards
$\alpha > 1$.
In DDH we show in different ways that, indeed,
no tidal transfer of mass from the satellite to the halo is expected
in a region where $\alpha \leq 1$. 
We address this point in simple cases using analytic approximations
in the impulse and adiabatic limits 
and then demonstrate the anticipated effects using merger N-body simulations.

\section{MASS-TRANSFER PRESCRIPTION AT \pmb{$\alpha > 1$} }
\label{sec:psi}

Once satellites continue to merge with a halo of an inner cusp $\alpha > 1$, 
the final slope is determined by the details of the tidal mass transfer from 
the satellite to the halo. We analyze this process via a quite general toy 
model prescription, which we justify and calibrate using N-body simulations.

\subsection{Toy Model Prescription}

We consider a halo of mean density profile $\bar\rho(r) \prop M(r)/r^3$ 
with a corresponding logarithmic slope profile $\alpha(r)$,
and a merging satellite of initial mass profile $m(\ell)$ with a
corresponding mean density profile 
$\bar\sigma(\ell) \prop m(\ell)/\ell^3$. 
Let $\mf(r)$ describe the final distribution of stripped satellite mass in 
spheres about the halo center. 
By equating $\mf(r)$ and $m(\ell)$ we obtain a one to one
correspondence between $\ell$ and $r$. We then define the initial-density ratio
\be
\psi[\alpha(r)] = {\bar\rho(r) \over \bar\sigma[\ell(r)]} \,.
\label{eq:psi}
\ee
This is an operational definition independent of any model assumption.
We argue below that $\psi(\alpha)$ is a relatively robust function
quite insensitive to the specific nature of the merger,
with the distinct property that it is a monotonically decreasing function
of $\alpha$, 
and with values significantly below unity in the range $\alpha >1$.
The robustness of $\psi(\alpha)$ is indicated in \se{simu} by a preliminary
set of merger simulations, to be confirmed and refined by a more
complete suite of simulations in an associated paper.

In order to obtain a qualitative feeling for why $\psi(\alpha)$ should
decrease with $\alpha$ we appeal to a very simplified toy model. 
We assume that the halo is spherical and fixed during the merger, 
and that the satellite is spherical and is tidally stripped shell by shell 
outside a momentary tidal radius.
The stripped material from a given satellite shell, identified by its original 
radius $\ell$, is assumed to be deposited on average in a halo radius $r$.

The condition $\psi=1$ refers to a crude resonance condition, 
$\bar\sigma(\ell)=\bar\rho(r)$, where the
orbital period at $\ell$ within the (unperturbed) satellite equals the orbital
period of the satellite within the halo at $r$. This is commonly assumed
to approximate the momentary tidal radius (e.g., Syer \& White 1998; Klypin
1999a). However, this condition ignores the inevitable $\alpha$ dependence 
of the stripping process, and it fails to address the structural changes
of the satellite before stripping and the difference between
where the stripping occurs and where the mass is actually deposited.

We demonstrated in DDH that the tidal mass transfer 
becomes weak at $\alpha <1$, which means that $\psi$ is higher there compared
to where $\alpha>1$.  If we define the tidal radius $\ellt$
by the Lagrangian point where the net force in the satellite rest frame
vanishes along the line connecting the centers of mass of halo and
satellite, we obtain for a satellite on a circular orbit
\be
\alpha(r)\, \bar\rho(r) = \bar\sigma(\ellt)
\label{eq:tidal_circ}
\ee
and for a satellite on a radial orbit
\be
[\alpha(r)-1]\, \bar\rho(r) = \bar\sigma(\ellt) \,.
\label{eq:tidal_rad}
\ee
If the satellite structure inside the tidal radius is assumed to remain
fixed, and if the stripped material outside $\ellt$ is assumed to be 
effectively deposited at the $r$ where it is
stripped, we obtain $\psi(\alpha) \propto \alpha^{-1}$ and 
$\propto (\alpha-1)^{-1}$ respectively. 
This is the kind of decrease in $\psi(\alpha)$ expected 
due to the $\alpha$ dependence of the stripping efficiency.

Another source of $\alpha$ dependence is due to the systematic
difference between the radii of stripping and deposit.
Along a typical orbit, the satellite distance from the halo center oscillates
periodically between the radii of apocenter $\rapo$ and pericenter $\rperi$,
while their amplitudes gradually decay due to dynamical friction.
Ghigna \etal (1998) studied the distribution of satellite orbits in a
high-resolution N-body simulation of a cluster emerging from a CDM cosmological
background and found that the median ratio $\rapo/\rperi$
is 6:1, with radial orbits common and circular orbits rare,
and with a distribution of eccentricities quite independent of $r$.
They also demonstrated that the tidal radii of the satellites
are consistent with being determined near pericenter, approximately under the
general resonance condition $\bar\rho(\rperi) \sim \bar\sigma(\ellt)$. 
Particles that escape from the satellite can be assumed,
on average, to continue on an orbit about the halo center with apocenter
and pericenter radii ``frozen" at their values near the time of escape,
suffering no further decay due to dynamical friction. Since a particle spends
most of its time near the apocenter of its orbit, we can say that the
escapers are effectively ``deposited" near the apocenter radius valid at
the time of escape. 
We can thus assume that the ratio of stripping radius ($\gsim \rperi$)
and deposit radius ($\lsim \rapo$) is typically
$\epsilon \gsim \rperi/\rapo \sim 1/6$.
If we ignore the structural evolution of the satellite inside
the tidal radius as well as the $\alpha$ dependence of the stripping
efficiency addressed above, we obtain straightforwardly a decrease as 
a function of $\alpha$ of the sort $\psi(\alpha) = \epsilon ^\alpha$,
where $\alpha$ is the effective slope of the mean density profile between
$\rperi$ and $\rapo$.

A third source of $\alpha$ dependence is the distortion of the satellite
inside the tidal radius, with an $\alpha$-dependent stretching along one axis
and an $\alpha$-independent compression along the others. 
The resultant decrease in satellite density before stripping
has been seen in simulations (e.g. Klypin \etal 1999a, Fig.~6; Hayashi \etal
2002).  At higher $\alpha$ values the stretching is stronger, which should 
make the stripping even more efficient there. 

The above three effects provide qualitative hints for the expected properties
of $\psi(\alpha)$, and in particular for it being a decreasing function
of $\alpha$ with values below unity. We turn to N-body simulations in order to
quantify these predictions.

\subsection{Merger Simulations}
\label{sec:simu}

In order to evaluate $\psi(\alpha)$ at $\alpha >1$ and test its robust 
properties,
we ran several N-body simulations of isolated mergers between a large halo and
a satellite halo of mass ratio $m/M=0.1$. Similar simulations were used in
DDH to study the inner core region and they are also described there
using several additional figures which are not essential for our current
purpose.
We use the Tree code by Mihos \& Hernquist (1996 and references
therein) but with dark-matter halos only (no gaseous disks).
In our default simulations, the large host halo is represented by $10^5$ 
equal-mass particles and the satellite by $10^4$ particles.
We re-ran one case with $0.55\times 10^6$ particles and the force resolution 
higher by a factor $\sqrt{5}$, and confirmed that the results were 
practically identical. 
The simulation units are: length $3.5$kpc, mass $5.6\times
10^{10}\msun$, and time $13.06$Myr. 
The force softening length is 0.08 units, i.e. $0.28$kpc

The initial halo density profile, as measured in the unperturbed initial
conditions, is a truncated isothermal sphere with a flat core,
\be
\rho(r) = {\rhoc\, e^{-(r/\rt)^2} \over 1 + (r/\rc)^2} \,,
\label{eq:halo_prof}
\ee
with $\rhoc=10.36$, $\rc=1$ and $\rt=10$
The density at the characteristic radius $\rc$ is thus $\rho(\rc)=5.13$.
The internal velocities are constructed to fulfill the isotropic Jeans
equation which ensures an equilibrium
configuration as discussed in Mihos \& Hernquist (1996).
When run in isolation, the halo profile has been tested to be very stable
for many dynamical times.
The logarithmic slope $\alpha(r)$ of the mean density profile
spans the range of interest between $\alpha=0$ and 3, and
its variation as a function of radius can be
described to a good approximation by $\alpha(r) \approx 1.73 \log r +0.67$
throughout the range $0.3 \leq \alpha \leq 2.9$. 
The initial halo density profile can be seen in \Fig{prof} below.
It resembles the generalized NFW profile
of \equ{nfw} with a core of $\alpha \ll 1$.
We chose to start with an inner core in order to also simulate the rapid
steepening to a cusp, but the outer profile, where $\alpha >1$, is quite 
generic.

The satellite initial density profile is fit by a Hernquist profile,
\be
\sigma(\ell) = {\sigc \over (\ell/\ellc)\, [1+(\ell/\ellc)]^3} \,,
\label{eq:sat_prof}
\ee
with the default choice $\sigc=19.2$ and $\ellc=1$ defining our
typical ``compact" satellite.
The initial satellite density profile can also be seen in
\Fig{prof}.
In the inner region it convergence to the NFW profile, $\ai=1$.
If we fit this Hernquist profile with an NFW profile by matching
the characteristic radii where the local logarithmic slope is $-2$,
we find that the radius corresponding to the characteristic
radius of the NFW profile is $\ell=\ellc/2$. At this radius the
density is $16\sigc/27$, which for our default satellite is
$\sigma(\ellc/2)=11.38$.

The satellite parameters were chosen to roughly mimic a typical 
compact satellite according to the distribution of halo properties
in the $\Lambda$CDM scenario.  In a hierarchical clustering scenario with a
fluctuation power spectrum $P_k \prop k^n$, the 
halo characteristic radii and densities scale like
$\ellc/\rc \propto m^{(1+\mnu)/3}$ and $\sigc/\rhoc \propto  m^{-\mnu}$,
with $\nu = (3+n)/2$. 
For $\Lambda$CDM on galactic scales one has $\mnu \simeq 0.33$.
This is confirmed by cosmological simulations (NFW; Bullock \etal 2001a),
where the typical halo profile is NFW.
The halos of lower masses are thus typically more compact; they
tend to have lower characteristic
radii and higher corresponding mean densities within these radii
(corresponding to higher virial concentration parameters).
For CDM halos of mass ratio $m/M=1/10$,
the typical ratio of characteristic radii and corresponding densities
are expected to be roughly $0.4$ and $2.1$.  
The corresponding ratios in the initial conditions of our simulations
are approximately $0.5$ and $2.2$, 
providing a reasonable match.
It may also be interesting to note that the satellite mean density interior
to $\ellc/2$ (or $\ellc$) is about $3.9$ (or $1.09$) times the halo mean 
density interior to $\rc$.

We simulated three cases of initial merger orbits:
a radial orbit, a circular orbit, and a typical elongated orbit with
$\rperi/\rapo\sim 1/6$ initially.
The unperturbed satellite is put initially at $r=20$, i.e.
at about $2\rt$, where the initial circular period is about $230$Myr.
In the circular and radial cases the magnitude of the initial satellite
velocity was set to
equal the circular velocity of the halo at that radius, namely a bound orbit
with an orbital kinetic energy that equals half the absolute value of
the total energy. 
For the elongated orbit the initial tangential velocity was one half of the
circular velocity at $r=20$.
Each merger has been followed until the satellite's bound core has practically
settled at the halo center. 

\begin{figure}[t]
\vskip 8.0cm
{\includegraphics{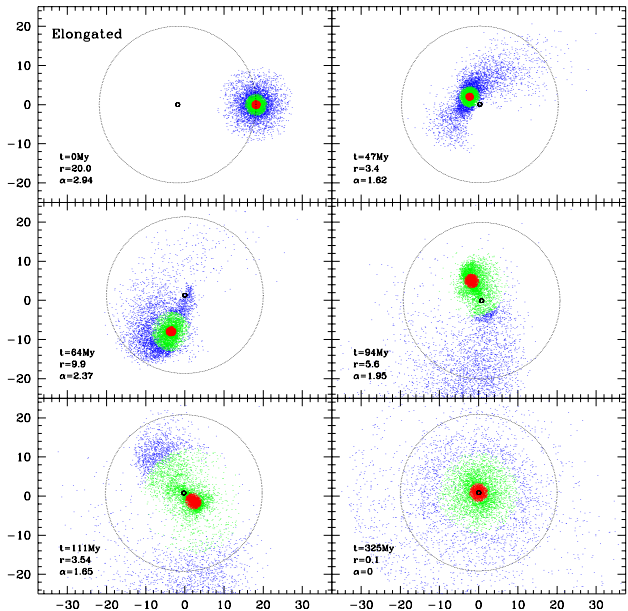}}
\caption{
The $10^4$ satellite particles in 6 snapshots during the elongated merger,
projected onto the orbital plane.
The $10^5$ live halo particles are not shown.
The center of mass is at the origin and the dot
marks the temporary halo maximum density.
The circle is of radius $r=20\simeq 2\rt$ about the halo maximum density,
corresponding to where the initial halo practically ends and where
the satellite is at the onset of the simulation.
The snapshots correspond to the initial conditions, the first pericenter,
the following three apocenters and the final distribution of stripped satellite
material.
The three thirds of the mass, in concentric shells about
the satellite bound center at each time, are marked by different colors.
The distance of the satellite center from the halo center is marked by $r$
(while radii within the satellite are marked by $\ell$).
}
\label{fig:snaps_elo}
\end{figure}

\Fig{snaps_elo} shows the satellite mass distribution in 6 snapshots
during the elongated merger, projected onto the orbital plane
(while DDH show analogous plots for the radial and circular mergers).
The satellite oscillates about the halo center through a sequence of pericenter
and apocenter passages.
The shape of the orbit remains roughly constant while it shrinks in scale; 
when measuring the ratio of pericenter to the following apocenter it is
$\rperi/\rapo \sim 1/3.5$, and when measuring the ratio of pericenter to the
preceding apocenter it is $\rperi/\rapo \sim 1/3.5$, but these ratios
remain roughly the same for all detectable pericenters.
The oscillations decay due to dynamical
friction until the satellite becomes confined to the halo core 
after $\sim 125$Myr and 5 pericenter passages.
By the first pericenter, the satellite is already stretched and stripped 
along its orbit, while it is temporarily shrunk in the perpendicular 
direction (as expected in DDH, see \se{tide}).
This is followed by a re-bounce and significant mass loss about the 
following apocenter. 
The visual impression confirms the notion that the particles that are 
torn away near a pericenter radius continue on orbits that reflect on average
the satellite orbit at the time of stripping, while the bound remnant 
continues to sink into smaller radii due to dynamical friction.
For example, in the snapshots corresponding to the second and third apocenters 
we clearly see a large amount of stripped satellite material spread about the
location of the previous apocenter.
We can crudely say that mass is stripped near pericenter and is practically 
``deposited" about the following apocenter radius.
The final distribution of satellite mass extends quite smoothly 
about the halo center in a puffy oblate ellipsoid that looks quite symmetric
in the orbital plane.

\begin{figure}[t]
\vskip 9cm
{\includegraphics{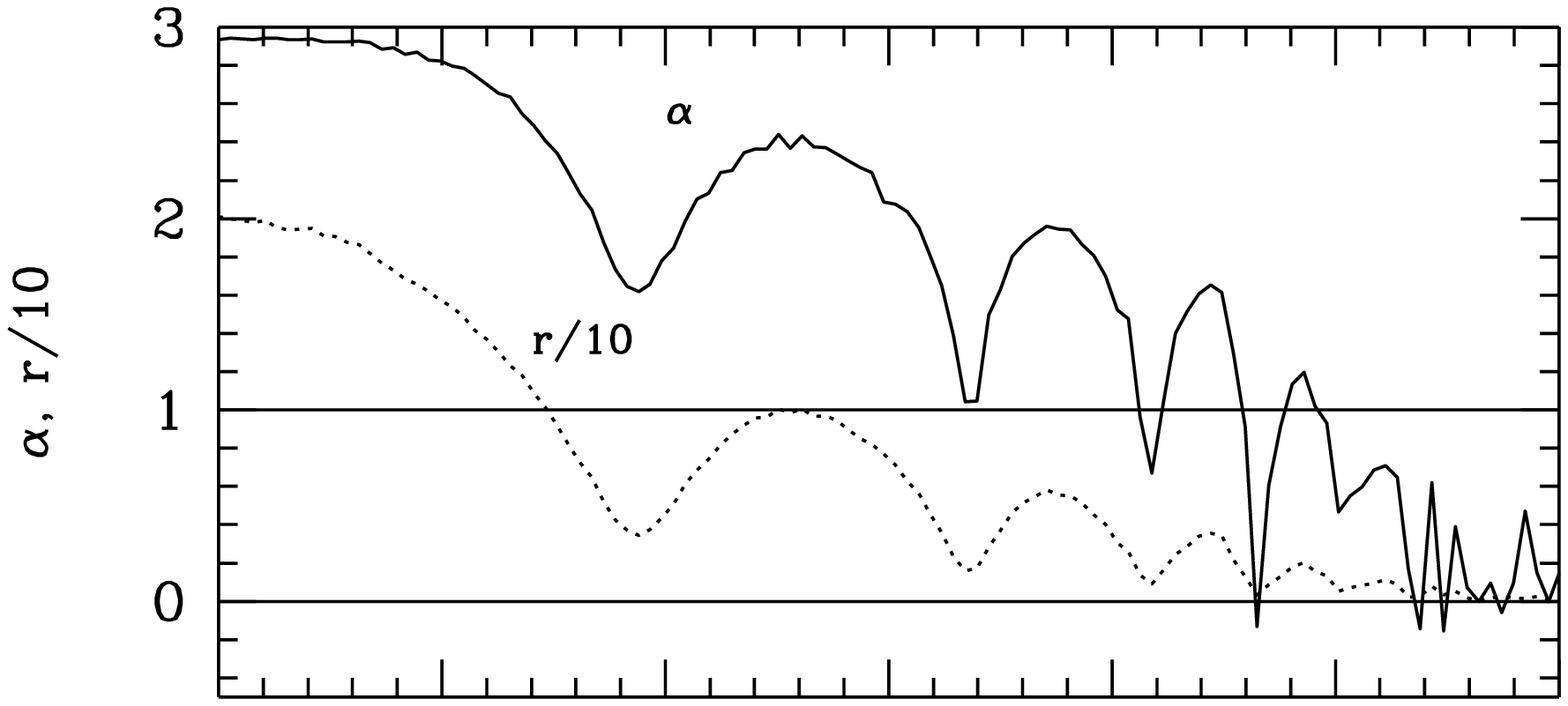}} 
{\includegraphics{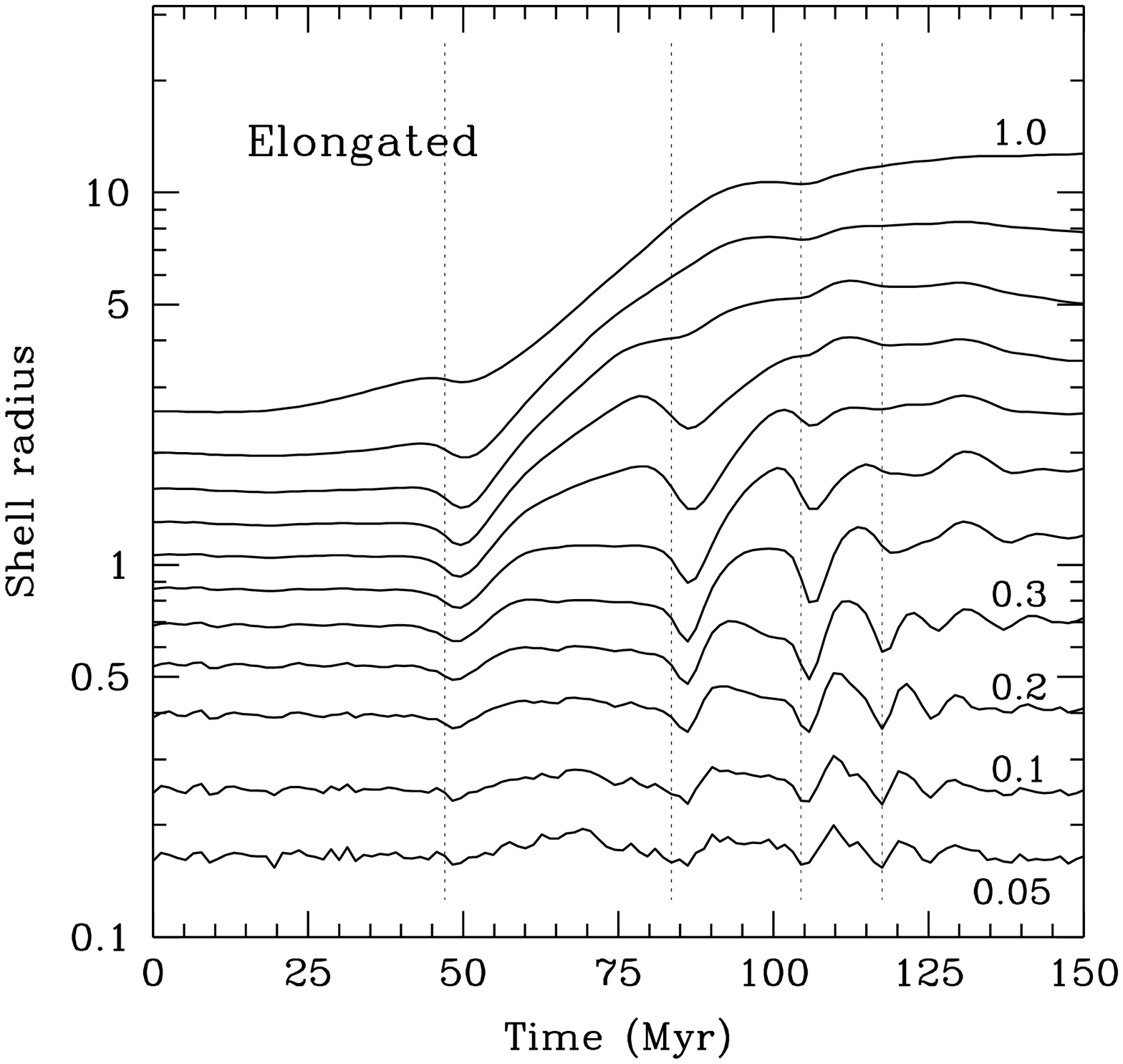}}
\caption{
Time evolution of the satellite spherical mass profile during the elongated
merger.
Shown in the bottom panels are the mean radii of concentric spherical shells
about the momentary satellite center, each encompassing a given
fraction of the original satellite mass, as marked on the right.
The top panels show the time evolution of position $r$ of the satellite
center relative to the halo center, and the
corresponding local slope $\alpha(r)$.
The times of pericenter passage are marked
in the bottom panel by vertical lines.
The stripping of a shell can be crudely identified by a rapid increase
of its radius.
Lack of stripping (and even a slight overall contraction of the innermost
bound shells) is noticed whenever the satellite enters the halo core.
}
\label{fig:radii}
\end{figure}

\Fig{radii} describes the time evolution of the satellite spherical
mass profile during the elongated merger by showing the mean radii of 
concentric spherical shells
about the satellite maximum-density center, each encompassing a given 
fraction of the satellite mass (i.e., not necessarily the same population of
particles at different times). 
Pericenter passages are identified five times.
Overall contraction of bound satellite shells seems to start roughly at 
these times, as expected both in the impulse and adiabatic limits (see DDH). 
Each major contraction is followed by a re-bounce as the 
satellite moves towards apocenter,
which results in overall expansion and stripping of the outer shells.
This seems like a manifestation of the expected delayed stripping due to the
energy pumped into the satellite by the inwards impulse in the core,  
as well as the adiabatic stretching and stripping outside the tidal radius.
Once the satellite becomes confined to the inner halo where $\alpha < 1$,
after $\sim125$Myr and 5 apocenters,
there is no apparent overall shell expansion anymore, indicating
that stripping has stopped, as expected.
Similar effects are noticed in DDH for the radial and circular mergers.

\begin{figure}[t]
\vskip 5.5cm
{\includegraphics{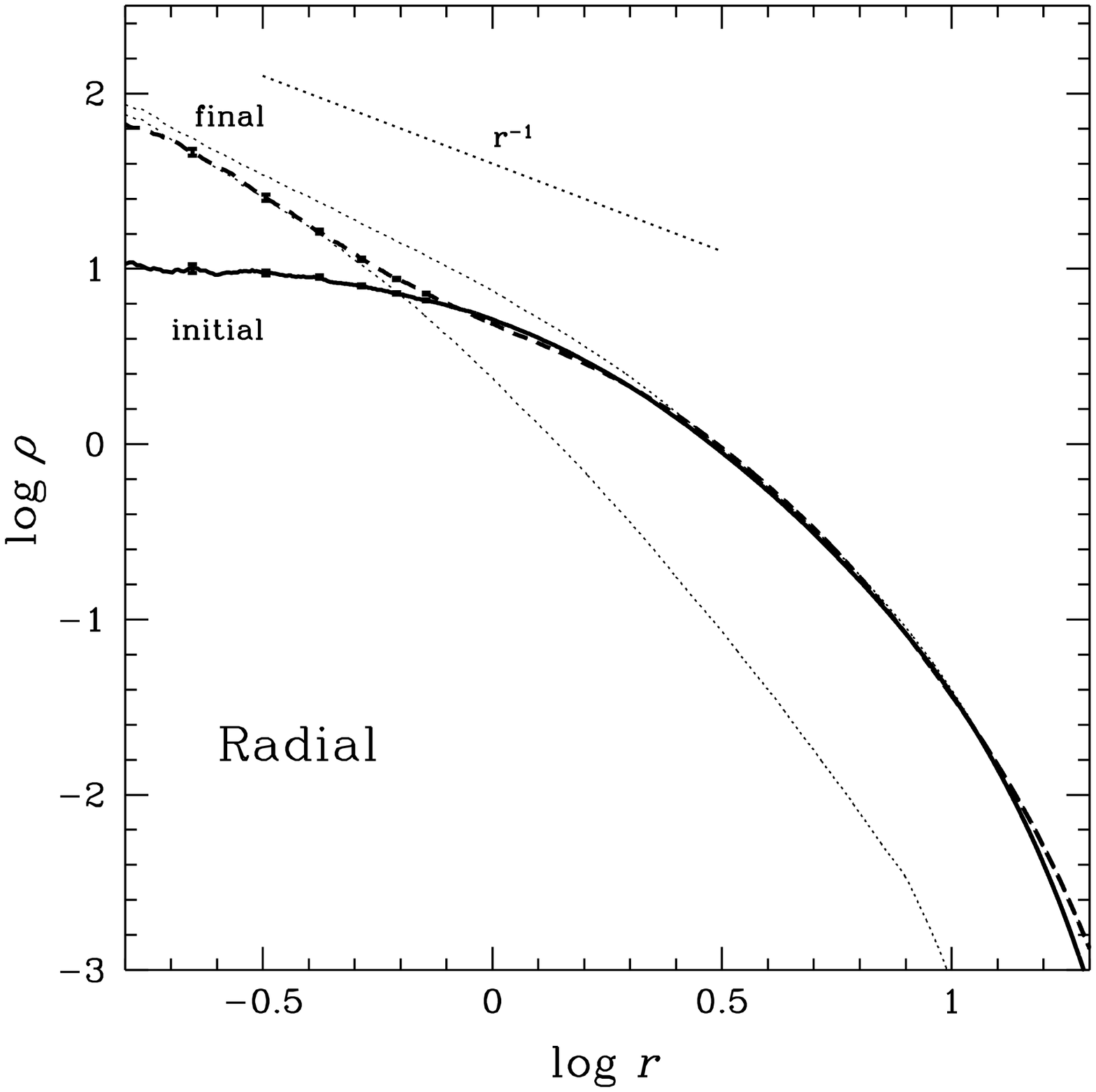}}
{\includegraphics{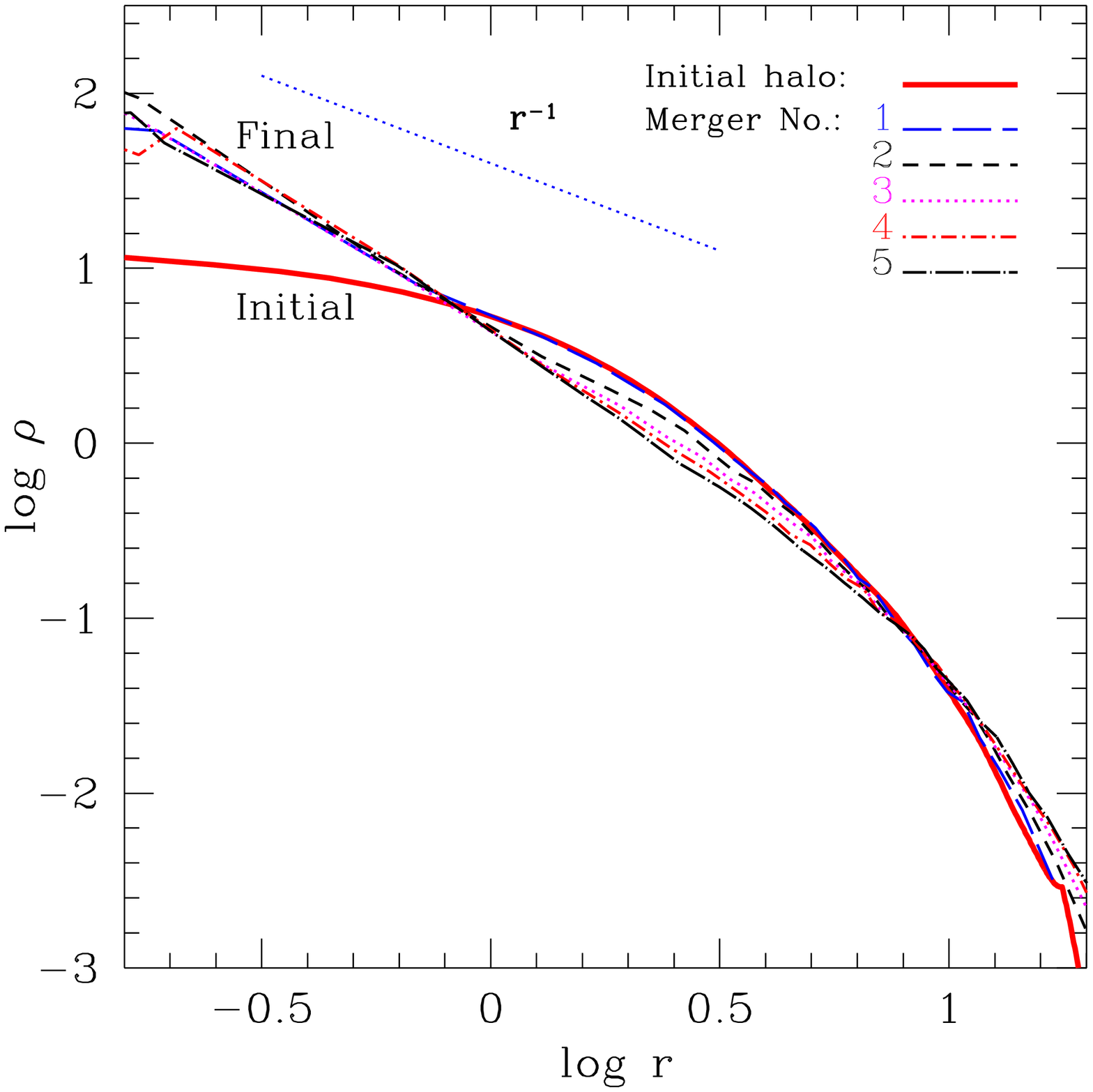}}
{\includegraphics{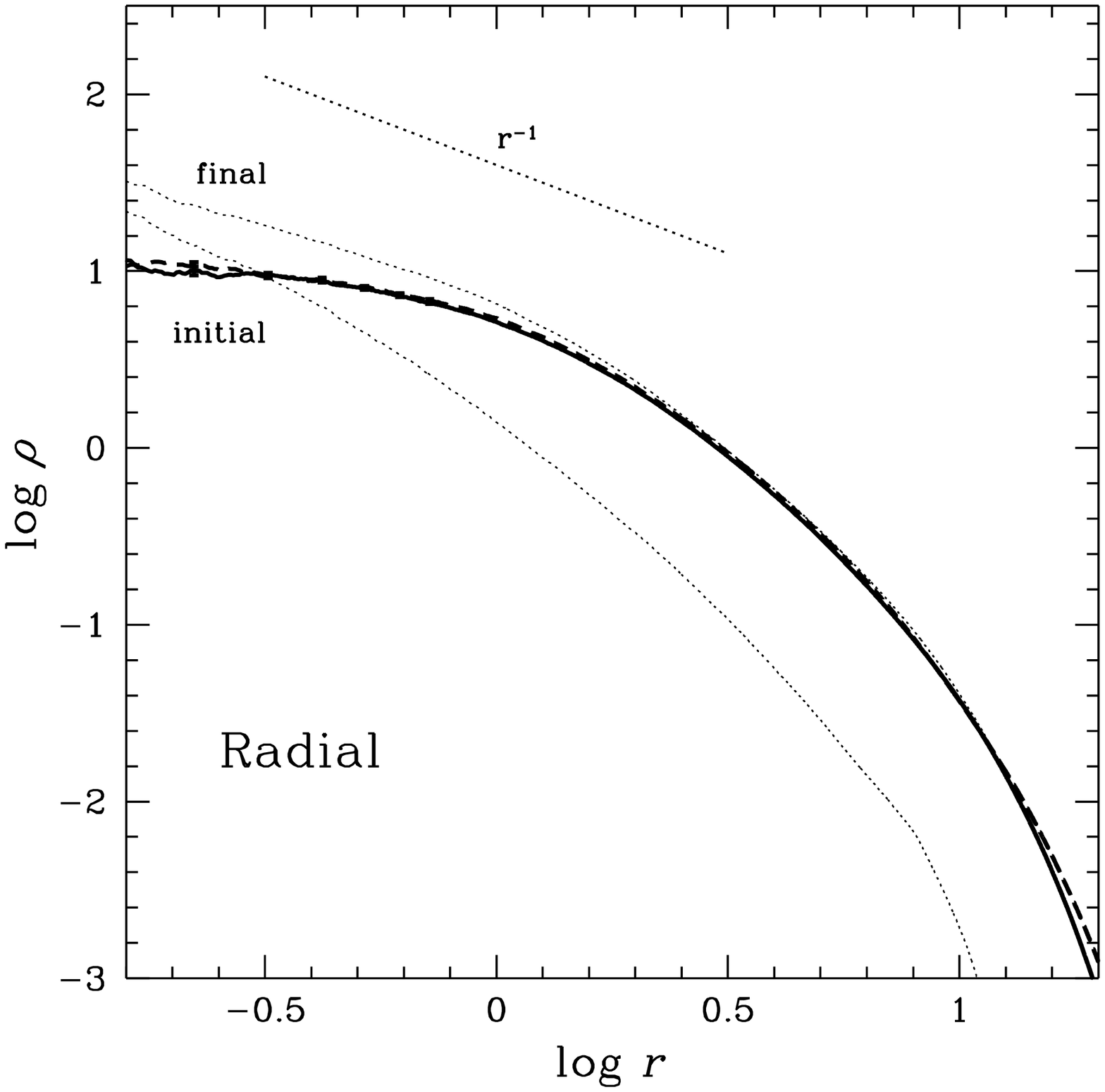}}
\caption{
Halo density profile before (solid) and after (dashed) the simulated mergers
with satellites of mass ratio 1:10. The initial halo has a flat core.
{\it Left:} one merger with a compact satellite.
A significant fraction of the satellite settles intact
at the halo center without depositing mass near $\alpha \lsim 1$,
causing the steepening of the core into a cusp.
{\it Middle:} tandem mergers with compact satellites on random elongated orbits.
After each merger, the total mass of the halo is scaled down back to
the original mass.  The cusp remains stable under the subsequent mergers.
{\it Right:} one merger with a puffy satellite, whose density has been 
scaled down by a factor of 3.3 compared to the compact satellite.
The puffy satellite loses most of its mass in the outer halo, leaving
the halo core practically unaffected.
}
\label{fig:prof}
\end{figure}

\Fig{prof} addresses the cusp formation straightforwardly
by showing the density profile of the halo before and after the
merger on a radial orbit. In DDH we also show similar plots for 
the circular and elongated merger orbits, with very similar results.
For the standard, compact satellite, 
the left panel demonstrates the inevitable steepening of the profile
in the core region, starting near the core boundary where $\alpha \simeq 1$.
The slight depletion of final density
in the region near $\alpha \sim 1$, compared to the slight increment in
halo density at larger radii, is consistent with no mass transfer in this
region while the orbit of the remaining satellite continues to decay into
smaller radii. 

How stable is this cusp under further mergers with similar satellites? 
To test this, we performed a series of mergers following each
other. The main progenitor halo in each merger
is taken to be the outcome of the previous
merger, except that its total mass is scaled down back to the original halo
mass (such that all the density profiles can be directly compared).
We achieve this by letting each simulation involve the same numbers of
$N$ halo particles and $n$ satellite particles (each of a fixed mass $m$),
where the halo particles are selected at random from the $N+n$ particles
of the halo produced in the previous merger.
In these tandem mergers we used $N=20,000$ and $n=2,000$.
The halo profile in the range of interest is found to remain stable
for at least several hundred Myr when the halo is run in isolation.
The satellite starts with the same profile as the original compact satellite.
It is put at the same
distance of $r=20$ as before but in an elongated orbit of a random spatial
orientation. The initial satellite velocity is tangential to the line
connecting the centers of mass of halo and satellite, with an amplitude 
in units of the circular velocity at $r=20$ 
chosen at random in the range $v/v_{\rm c}=0.2-0.6$ 
(compared to the typical case simulated before, where $v/v_{\rm c}=0.49$
led to peri/apocenter ratio of roughly 1:6).
Each merger was followed for 325Myr before the following merger started.
The density profiles after each of the first 5 mergers are shown
in the middle panel of \Fig{prof}. 
The first merger reproduces a cusp very similar to the cusp produced
when the merger was simulated with $N=100,000$ (and with $N=500,000$),
indicating that even $N=20,000$ is adequate for crude results in the regime
of interest here.
We see that inside the initial core radius the
cusp is stable at a slope of $\alpha \simeq 1.5$,  
with no systematic tendency to deviate
from the power-law profile resulting already after the first merger.
At the radius where $\alpha \gsim 1.5$ the profile flattens, and 
the amplitude becomes
lower partly because of the renormalization of the total mass between
every two mergers.
Our main goal in the current paper is to try to understand the origin of 
the stable cusp as indicated by these simulations of tandem mergers.

How much puffing is needed in order to prevent damage by merging satellites
to the halo core?
In order to obtain a first clue, we have performed simulations
similar to those described above,
except that the compact satellites of CDM have been replaced by more puffy
satellites of the same total mass.
In the Hernquist profile, \equ{sat_prof}, where the default compact satellites
had $\sigc=19.2$ and $\ellc=1$, the parameters of
the puffy satellites are now $\sigc'=\sigc/8$ and $\ellc'=2\ellc$.
With this choice, the density at $\ell=1/2$, the characteristic radius
of the original compact satellite, is scaled down by a factor of $2.3$,
to just below the density of the host halo at its characteristic radius.
This corresponds to a reduction by a factor of $2.8$ in the
mean density interior to $\ell=1/2$.
The right panel of \Fig{prof} shows the effect of such a merger
with a puffy satellite
on the halo density profile in the case of a radial radial merger
(DDH also show similar results for other merger orbits).
When the satellite is puffy, we see that
almost all the satellite mass is stripped before the satellite
orbit decays to the $\alpha \leq 1$ zone, and as a result the halo core
is practically unaffected. We learn that a modest reduction in the
initial satellite inner density is enough for preventing the cusp
formation seen in the compact-satellite case.

\subsection{Measuring \pmb{$\psi(\alpha)$}\ in the Simulations}

The measurement of $\psi(\alpha)$ in the simulations is straightforward
(and free of any model assumption).
We simply measure the mass profile of the stripped satellite mass about the
halo center at the final time, $\mf(r)$, and equate it with the initial
satellite mass profile $m(\ell)$ to obtain the deposit relation $\ell(r)$. 
Then $\psi$ is evaluated from \equ{psi} at any desired $r$, and is expressed
as $\psi(\alpha)$ given the halo slope profile $\alpha(r)$.

\begin{figure}[t]
\vskip 8cm
{\includegraphics{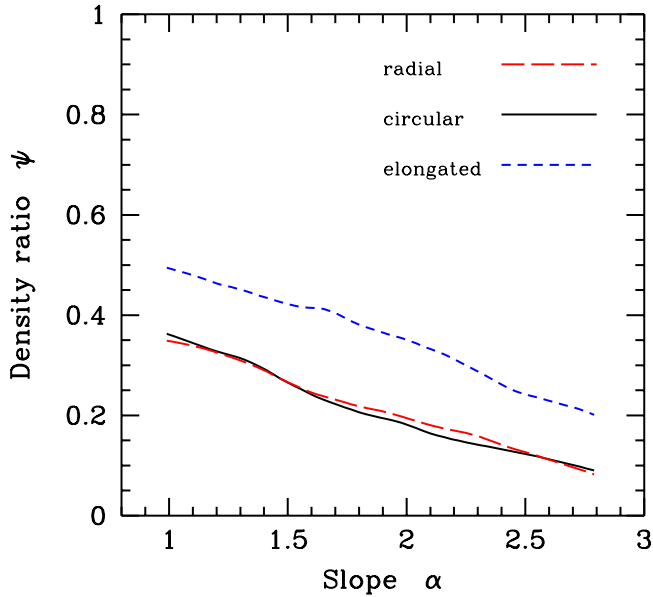}}
\caption{
The mass-transfer condition.
Density ratio $\psi$, at deposit, as a function of halo slope $\alpha$.
Shown are the measured results from the three different N-body merger
simulations. A qualitative general fit in the range $\alpha>1$
is provided by the power law $\psi(\alpha) \simeq 0.5/\alpha$,
or by the exponential $\psi(\alpha) \simeq 0.5^\alpha$.
}
\label{fig:psial}
\end{figure}

\Fig{psial} shows the measured $\psi(\alpha)$ in three different
merger simulations, where the orbits are very different from each other:
circular, radial and elongated 1:6.
All the curves show a similar general behavior, consistent with our qualitative
expectations.
First, the value of $\psi$ is significantly below unity, demonstrating that 
the mass transfer is more efficient than predicted by the naive $\psi=1$ model.
Second, the function $\psi(\alpha)$ is monotonically decreasing with increasing
$\alpha$ in the range $1<\alpha<3$. The general behavior can be crudely
approximated by several different functions, such as the power law 
$\psi(\alpha) \simeq 0.5/\alpha$, 
or the exponential $\psi(\alpha) \simeq 0.5^\alpha$.
The scatter about these fits is roughly $\pm 0.1$.
The relative robustness of $\psi(\alpha)$ to the nature of the merger orbit
indicates that \equ{psi} may serve 
as a useful approximate recipe for tidal mass transfer in a general merger.

In order to evaluate the sensitivity of our results to the resolution of
the simulation, we ran a case identical to the merger on an elongated orbit
described above, but now with 5 times more particles ($N=0.55\times 10^6$)
and a softening length smaller by a factor of $5^{1/3}$ accordingly.
The results are found to be practically identical in all respects,
namely the decay rate of the satellite radius within the halo,
the mass loss from the satellite, the final halo density profile,
and the derived $\psi(\alpha)$.
This indicates that the resolution of our simulations is adequate
in the range of radii of relevance, well inside the core/cusp
and down to below $0.2\rs$.

In the current simulations we have only tested in a limited way
the relative robustness to the merger orbit.
The simulated halo and satellite are of a typical mass ratio and their profiles
relate to each other in general accordance with the expected average
scaling in the $\Lambda$CDM cosmology, which makes the obtained 
$\psi(\alpha)$ a sensible first guess. However, the robustness of
$\psi(\alpha)$ to variations about this mass ratio and these profiles  
should be confirmed and refined using a more complete suite of merger
simulations. We adopt below $\psi(\alpha)=0.5/\alpha$ as an illustrative
example for obtaining numerical results but the following analysis is 
valid for a quite general $\psi(\alpha)$.

\section{AN ASYMPTOTIC PROFILE: QUALITATIVE}
\label{sec:qual}

We saw in DDH that when the halo profile is flat, $\alpha \leq 1$,
tidal compression causes rapid steepening to $\alpha > 1$.
In this section we investigate the development of the profile due to a sequence
of mergers between similar halos where $\alpha$ is of order unity
or larger.
We show that if the tidal stripping is described by a condition similar
to \equ{psi}, with $\psi(\alpha)$ monotonically decreasing 
rapidly enough, then the profile evolves slowly towards 
an asymptotic stable power law $r^{-\aas}$, with $\aas$ larger 
than unity, which we can evaluate once $\psi(\alpha)$ is given. 
The evolution to this asymptotic profile is through a sequence
of profiles which crudely resemble the generalized NFW shape,
and with an inner cusp of slope slightly larger than unity, $1<\alpha<\aas$.
While the value of $\aas$ may depend on the exact shape of $\psi(\alpha)$,
the following analysis showing the convergence to an asymptotic profile is not 
sensitive to it as long as $\psi(\alpha)$ is decreasing rapidly enough.

\subsection{Homologous Halos}

We consider the buildup of the halo of mass $M$ by a sequence of mergers
with satellites of masses $m\leq M$ drawn from a cosmological distribution of 
halos (similar to Syre \& White 1998).
In our search for a self-similar evolution, 
we assume that the halo and satellite before the encounter are 
{\it homologous}, meaning that their unperturbed mean density  
profiles are scaled version of each other,
\be
\bar\sigma(\ell) = \varrho\, \bar\rho(\ell/\lambda) \,,
\quad
\varrho = m/\lambda^3\,,
\label{eq:homo}
\ee
where $m$ stands for the satellite-halo mass ratio $m/M$ (we use hereafter
$M=1$).
If we refer, for example, to the functional form of \equ{nfw}, then
$\varrho=\sigc/\rhoc$ and $\lambda=\ellc/\rc$.
The cosmological N-body simulations of CDM as well as of power-law power
spectra (NFW; Bullock \etal 2001a) 
show that halos of lower masses tend to
have lower characteristic
radii and higher corresponding mean densities within these radii
(corresponding to higher virial concentration parameters).
The scaling of the averages of these quantities, at any given time, 
is found to be
\be
\lambda = m^{(1+\mnu)/3} \,,
\quad
\varrho =  m^{-\mnu} \,,
\label{eq:mnu}
\ee
with $\mnu \simeq 0.33$ for $\Lambda$CDM.
Given a power-law power spectrum $P(k) \propto k^n$, a simple scaling
argument based on linear theory predicts this behavior with $\mnu =(3+n)/2$.
Thus, $\mnu \simeq 0.33$ corresponds to $n \simeq -2.3$, as expected
for a $\Lambda$CDM power spectrum at the relevant scales.

\subsection{Convergence to an Asymptotic Profile}

\begin{figure}[t]
\vskip 6.0 truecm
\includegraphics{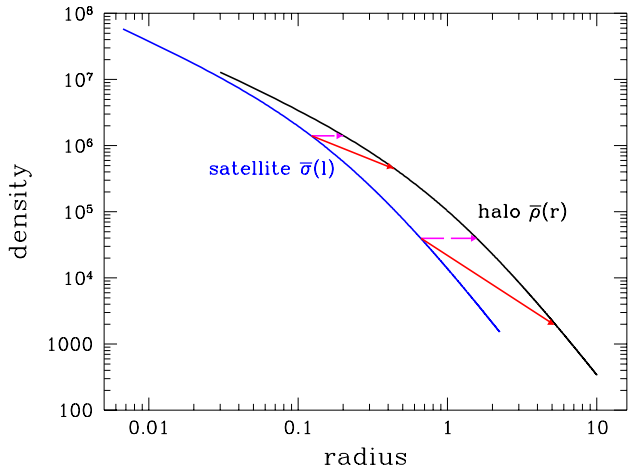}
\caption{ 
A schematic illustration of satellite mass deposit in the halo.
Shown are an NFW halo profile $\bar\rho(r)$, and a homologous satellite
profile $\bar\sigma(\ell)$ properly shifted to the left and upwards.
The horizontal dashed arrows refer to deposit based on $\psi(\alpha)=1$.
This would steepen the profile, as steep regions of 
$\bar\sigma(\ell)$ are deposited at flatter regions of $\bar\rho(r)$.
The solid arrows illustrate realistic tidal mass transfer, with a decreasing
$\psi(\alpha)$.
The vertical displacements, which grow with $r$, refer to $\psi(\alpha)<1$.
In this case, the slope at $\ell$ is closer or even flatter than the slope
at $r$, and more mass tends to be deposited at large $r$,
leading to flattening of the halo profile.
}
\label{fig:intuitive}
\end{figure}

The origin of an asymptotic slope 
can be qualitatively understood in simple terms
via the illustration in \Fig{intuitive}. We show a schematic mean density
profile of a halo and a homologous satellite scaled accordingly.
As in the toy-model interpretation of \equ{psi}, we assume that satellite 
shell $\ell$ is being deposited at halo radius $r$ where the slope is 
$\alpha$, with a given decreasing function $\psi(\alpha)$.
To visualize this deposit process,
we connect by arrows two points on the log-log curve $\bar\sigma(\ell)$ with
the corresponding points on the log-log curve $\bar\rho(r)$, illustrating both
the (horizontal) distance ratio $r/\ell$ and the (vertical)
density ratio $\psi$.
The simplified condition $\psi=1$ (as adopted, e.g., by Syre \& White 1998)
leads to a correspondence via horizontal arrows.
In this case, the satellite slope at $\ell$ is always
steeper than the halo slope at $r$, naturally leading to a steepening of the
halo profile. The mean profile evolves in this case towards the steepest 
possible power law, $\alpha=3$.
On the other hand, when $\psi$ is properly decreasing with $\alpha$,
the relation between $\ell$ and $r$ is now illustrated by the solid arrows,
where the vertical displacement is a growing function of $r$ and $\alpha$.
This is expected to weaken the steepening and possibly turn it into
flattening of the profile, because the slope at $\ell$ is now closer to the 
slope at $r$, and may be even flatter. Another way to understand this
flattening effect is by noticing that due to the decrease of $\psi(\alpha)$ 
more satellite mass is deposited at outer halo radii.

We can be a bit more quantitative as follows.
With $m \ll M$, we assume that the original halo is not
affected by tides, and that its mean density profile  
evolves only by the addition of the satellite mass wherever it is 
being deposited, namely,
\be
\bar\rho_{\rm final}(r) = \bar\rho(r) +\bar\sigma[\ell(r)]
{\ell(r)^3\over r^3} \,,
\label{eq:rho_final}
\ee
where $\ell(r)$ is defined by $\mf(r)=m(\ell)$.
Given that $\alpha(r)$ is minus the logarithmic derivative of $\bar\rho(r)$,
it is straightforward to show that the change of $\alpha$ at $r$ due to the
merger is
\be
\Delta \alpha(r) 
= -{ F(r)\over 1 + F(r)}\,
{\dd F(r) \over \dd r} \,, 
\quad
F(r) \equiv {\bar\sigma(\ell) \over \bar\rho(r)} {\ell^3 \over r^3} \,.
\label{eq:dalpha}
\ee
Note that the first term in $F(r)$ is $1/\psi(r)$,
and that $F(r) = m(\ell)/M(r)$.
 
Every power law is a self-similar solution, where $\Delta\alpha(r)=0$.
This is because when $\alpha(r)$ is constant, so is $\psi(r)$, namely
$\bar\sigma(\ell)/\bar\rho(r)$ is constant, implying that 
$\ell/r={\rm const.}$,
and therefore the derivative $\dd F/\dd r$ vanishes. 
However, $\Delta\alpha(r)=0$ does
not guarantee a stable solution to which the process would converge.

When $\alpha$ is varying with $r$,
as long as $\alpha$ is small and $\psi$ is close to unity,
the ratio $\ell/r$ is a decreasing function of $r$. 
Thus, if $\psi[\alpha(r)]=$const., 
the $\ell^3/r^3$ term determines a positive $\Delta\alpha(r)$, 
namely continuous steepening towards $\alpha=3$.
On the other hand, when $\psi(\alpha)$ is properly decreasing with $\alpha$, 
$1/\psi[\alpha(r)]$ is an increasing function 
of $r$, an increase which can balance the decrease of $\ell^3/r^3$ 
and produce a fixed point of $\Delta\alpha=0$ at some asymptotic slope $\aas$. 
This is a stable solution, where
$\Delta\alpha>0$ at $\alpha<\aas$, and $\Delta\alpha<0$ at $\alpha>\aas$.

\begin{figure}[t]
\vskip 6.7cm
{\includegraphics{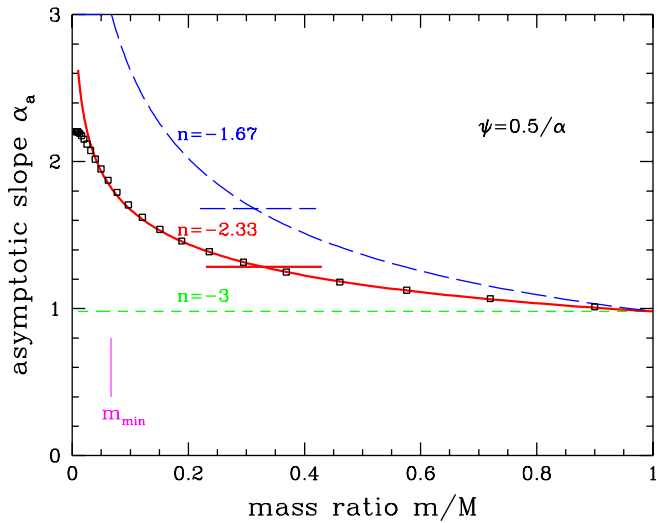}}
\caption{ 
The asymptotic slope $\aas$ due to a sequence of mergers with a
fixed mass ratio $m/M$. The solid curve is the prediction of linear theory,
\equ{aas}, for $\Lambda$CDM ($n=-2.33$). The symbols represent the asymptotic
results of the non-linear non-parametric
toy simulations (\se{toy}), demonstrating the success of the linear
approximation in the relevant range of $m/M$. 
The assumed mass-transfer model in both cases is $\psi(\alpha)=0.5/\alpha$.
The linear results for two other power spectra are shown as dashed curves
with $n$ marked.
The values of $\aas$ obtained by a distribution of satellite
masses [\se{dist}, \equ{a_aas_m}] are marked by the horizontal bars.
The minimum mass ratio imposed by the dynamical-friction time scale
[\se{dist}, \equ{tfric}] is indicated by the vertical bar.
}
\label{fig:aas}
\end{figure}

\subsection{Linear Perturbation Analysis: Summary}
\label{sec:linear_summary}

A rigorous perturbation analysis, for a given $\psi(\alpha)$ and $m/M$,
is described in the following section. We provide here a summary of this
analysis and its main result so that the reader who is not interested in the
mathematical details can skip \se{linear}, which is rather technical.

The linear analysis is based on the following two 
assumptions, which are argued to be of quite general validity: 
\begin{itemize}
\item 
Near the asymptotic solution the slope $\alpha(r)$ is varying slowly,
namely, the first and higher derivatives of $\alpha(r)$ are much smaller 
than $\alpha(r)$.  This allows us to write
$\alpha(r) \simeq \a0 + \epsilon \delta(r)$,
and examine the behavior of $\Delta\alpha$ at first order in the 
small parameter $\epsilon$.
\item 
The (very small) varying part of $\alpha(r)$ is approximately linear 
in $\ln r$, namely, the second and higher derivatives of $\alpha(\ln r)$ 
are much smaller than $\dd \alpha/ \dd \ln r$.
\end{itemize}
This analysis leads to an explicit expression for $\Delta\alpha$ at $r$:
\begin{eqnarray}
\Delta\alpha(r) &\simeq&
{\alpha'(r) r \over 1 
  + \lambda^{-3} \varrho^{-3/\alpha} \psi(\alpha)^{1-3/\alpha} 
 }\\
&& \times \frac{1}{\alpha^2}
   \left( \alpha(\alpha - 3)\frac{\psi'(\alpha)}{\psi(\alpha)}
   + 3 \ln[\varrho \psi(\alpha)] \right) \,.   \nonumber
\label{eq:dalpha_l}
\end{eqnarray}
Here $\alpha=\alpha(r)$, $\alpha'(r)\equiv \dd\alpha/\dd r$ and
$\psi'(\alpha) \equiv \dd\psi/\dd \alpha$.
Thus, the sign of $\Delta\alpha(r)$ is the sign of the expression
in big brackets, and the asymptotic slope $\aas$ is the 
solution of the equation
\be
\alpha(\alpha - 3)\frac{\psi'(\alpha)}{\psi(\alpha)}
+ 3 \ln[\varrho \psi(\alpha)]  = 0 \,.
\label{eq:aas}
\ee
The factor in front of the big parentheses in \equ{dalpha_l}
determines the rate of convergence to $\aas$;
note that the rate slows down as $\alpha$ approaches a constant and
$\alpha'\rightarrow 0$.
Recall that $\psi(\alpha)$ is given,
and that the factors $\varrho$ and $\lambda$ are functions of the mass 
ratio in the merger and the cosmological power spectrum, \equ{mnu}.

The lines in \Fig{aas} show the asymptotic slope $\aas$ as predicted by 
the linear theory, \equ{aas}, for a sequence of mergers with the same mass
ratio $m/M$, as a function of $m/M$. 
The assumed mass-transfer recipe in both cases is $\psi(\alpha)=0.5/\alpha$. 
The middle curve is for $\Lambda$CDM, $n=-2.33$, and 
the other curves are for $n=-3$ and $n=-1.67$.

The discussion so far referred to a sequence of mergers with the same mass
ratio. In \se{dist} we will generalize the analysis to the case of a realistic
cosmological distribution of mass ratios.

\section{LINEAR ANALYSIS OF THE ASYMPTOTIC PROFILE}
\label{sec:linear}

This section is rather technical, and can be skipped by the non-practitioner
reader without hurting the flow of the discussion.

\subsection{The General Case}

The assumptions and analysis become more transparent if we switch to
logarithmic variables,
\begin{equation}
\tilde{r} \equiv \ln r \,,
\quad  \tilde{\rho} \equiv \ln\bar \rho \ .
\label{eq:a_log}
\end{equation}
Hereafter we use a tilde to mark logarithmic quantities in general.
Then \equ{rho_final} becomes
\be
\tilde\rho_{\rm final}(\tilde{r}) =
\tilde{\rho}(\tilde{r}) +
  \ln [1 + e^{\tilde{\sigma}(\tilde{\ell}) - \tilde{\rho}(\tilde{r})
            + 3(\tilde{\ell} - \tilde{r}) } ] \, .
\label{eq:a_Lmerger}
\ee
Using the log variables,
the local slope is simply (minus) the first derivative of the density
profile,
\be
  \alpha(\tilde{r}) = -{\dd \tilde{\rho}(\tilde{r})}/{\dd\tilde{r}} \, .
\ee
Then, from \equ{a_Lmerger}, the change in the slope in the merger,
$\alpha_{\rm final}(\tilde r)-\alpha(\tilde r)$, is
\begin{eqnarray}
\Delta\alpha(\tilde{r}) &=&
 -\frac{1}{ 1 + e^{\tilde{\rho}(\tilde{r}) -
\tilde{\sigma}(\tilde{\ell})
             + 3(\tilde{r} - \tilde{\ell}) }} \,
 \frac{\dd}{\dd\tilde{r}} [ \tilde{\sigma}(\tilde{\ell})
    - \tilde{\rho}(\tilde{r}) + 3(\tilde{\ell} - \tilde{r}) ] \nonumber
     \\ &=&
  \frac{1}{ 1 + e^{\tilde{\rho}(\tilde{r}) -
\tilde{\sigma}(\tilde{\ell})
             + 3(\tilde{r} - \tilde{\ell}) }}
     \left[ \alpha(\tilde{\ell} - \tilde{\lambda})
        \frac{d\tilde{\ell}}{d\tilde{r}}
    - \alpha(\tilde{r}) - 3\left(\frac{d\tilde{\ell}}{d\tilde{r}}
-1\right)
      \right] \ .
\label{eq:a_new-alpha}
\end{eqnarray}
In the second equation we have assumed that the profiles of the halo and
satellite are {\it homologous}, \equ{homo}.

In order to make the above expression for $\Delta\alpha$ useful,
we need to know the function $\tilde\ell(\tilde r)$,
and especially its derivative $\dd \tilde\ell / \dd\tilde r$.
This is provided by the mass-transfer prescription, \equ{psi}, which in
the log variables takes the form
\be
\label{eq:a_Lstripping}
\tilde{\rho}(\tilde{r}) - \tilde{\sigma}(\tilde{\ell}) =
      \tilde\psi(\alpha) \,,
\ee
with the function $\psi(\alpha)$ given, and $\alpha=\alpha(\tilde r)$.

Substituting in \equ{a_Lstripping} the scaling relation for {\it homologous}
halo and satellite, \equ{homo},
we obtain the equation for mass transfer in the homologous case:
\begin{equation}
  \tilde{\rho}(\tilde{r}) - \tilde{\rho}(\tilde{\ell}-\tilde{\lambda}) =
     \tilde\varrho + \tilde\psi(\alpha) \,.
\label{eq:a_Lstripping_homo}
\end{equation}
Once the scaling factors $\lambda$ and $\varrho$ are given,
and the functions $\psi(\alpha)$ and $\alpha(r)$ are known,
this equation should allow us to evaluate the desired
$\tilde\ell(\tilde r)$, to be used in \equ{a_new-alpha}.

\subsection{First-Order Approximations}

For any pure power law, $\alpha(\tilde{r}) ={\rm const.}$, 
we have $\Delta\alpha(\tilde r)=0$ for all $\tilde{r}$'s.
This is because then $\psi$ is constant, meaning that
$\tilde\rho-\tilde\sigma$ is constant, 
so $\tilde\ell-\tilde r$ is constant, and then the
derivative vanishes in \equ{a_new-alpha}.
However, not all these solutions are stable. In order to find a stable
solution, we examine $\Delta\alpha(\tilde{r})$ for small perturbations
of order $\epsilon$ about $\alpha=\a0$, and see how the
first order in $\epsilon$ behaves.

We first wish to evaluate $\dd \tilde\ell / \dd\tilde r$ at first order
in
$\epsilon$.
Based on the definition of $\alpha$, the homologous mass-transfer condition,
\equ{a_Lstripping_homo}, can be written as
\begin{equation}
  \int_{\tilde{r}}^{\tilde{\ell}-\tilde{\lambda}} \!\! \dd s\, \alpha(s)
= \tilde\varrho +\tilde\psi(\alpha) \,.
\end{equation}
Into this equation we substitute the perturbed quantities
\begin{eqnarray}
\label{eq:epsi}
  \alpha(\tilde{r}) &=& \a0 + \epsilon\delta(\tilde{r}) \ , \\
  \tilde{\ell}(\tilde{r}) &=& \tilde{\ell}_0(\tilde{r}) 
       + \epsilon\tilde{\ell}_1(\tilde{r}) \ ,
\end{eqnarray}
where $\epsilon$ is small, $\delta$ is of the order of $\alpha$,
and $\ell_1$ is of the order of $\ell$. We now
collect terms of the same order.
The zero's order term yields
\begin{equation}
  [\tilde{\ell}_0(\tilde{r}) -\tilde{\lambda}- \tilde{r}] \a0 
 = \tilde\varrho +\tilde\psi(\alpha) \,,
\end{equation}
\begin{equation}
\label{eq:a_l0}
  \tilde{\ell}_0(\tilde{r}) 
 = [\tilde\varrho +\tilde\psi(\a0)] /\a0 \,
           +\tilde{\lambda} + \tilde{r} \, .
\end{equation}
The first order term gives
\begin{equation}
  \int_{\tilde{r}}^{\tilde{\ell}_0-\tilde{\lambda}} \!\! \dd s\,
\delta(s)
   + \tilde{\ell}_1(\tilde{r})\a0 
     = \frac{\psi'(\a0)}{\psi(\a0)}
        \delta(\tilde{r}) \ ,
\end{equation}
\begin{equation}
  \tilde{\ell}_1(\tilde{r}) = \frac{1}{\a0}
    \left[ \frac{\psi'(\alpha_0)}{\psi(\a0)}
        \delta(\tilde{r})
        - \int_{\tilde{r}}^{\tilde{\ell}_0 - \tilde{\lambda}} 
        \!\! \dd s\, \delta(s) \right] \,,
\end{equation}
where $\psi'(\alpha)\equiv \dd \psi / \dd \alpha$.
Using these results we can now calculate
${\dd\tilde{\ell}(\tilde{r}) /\dd\tilde{r}}$ to first order:
\begin{equation}
  \frac{d\tilde{\ell}(\tilde{r})}{d\tilde{r}} = 1 
    + \frac{\epsilon}{\a0} 
        \left[\frac{\psi'(\a0)}{\psi(\alpha_0)}\delta'(\tilde{r})
        + \delta(\tilde{r}) - \delta(\tilde{\ell}_0 - \tilde{\lambda})
    \right] \, ,
\end{equation}
where $\delta'(\tilde r) \equiv \dd \delta / \dd \ln r$.

When we plug this result into \equ{a_new-alpha}, we obtain after some
algebra
\begin{eqnarray}
\label{eq:a_first-order}
  \Delta \alpha(\tilde{r}) &=&  
    \frac{\epsilon}{ 1 + e^{\tilde{\rho}(\tilde{r}) 
                         - \tilde{\sigma}(\tilde{\ell}_0)
             + 3(\tilde{r} - \tilde{\ell}_0) }} \\
    &&\times \left(\frac{\psi'(\a0)}{\psi(\a0)}\delta'(\tilde{r})
        - \frac{3}{\a0}
        \left[\frac{\psi'(\a0)}{\psi(\a0)}\delta'(\tilde{r})
        + \delta(\tilde{r}) - \delta(\tilde{\ell}_0 -
\tilde{\lambda})\right]
     \right) + {\mathcal{O}}(\epsilon^2)\ . \nonumber
\end{eqnarray}
We can further simplify the above expression by substituting $\ell_0$
with
its explicit value taken from \equ{a_l0}. Using \equ{a_Lstripping_homo}
and the definition of $\lambda$, we obtain
\begin{eqnarray}
\label{eq:pre-dalpha}
  \Delta \alpha(\tilde{r}) &=&  
    \frac{\epsilon}{ 1 +
       \lambda^{-3} \varrho^{-3/\a0} \psi(\a0)^{1-3/\a0} 
     }\\
    &&\times \left\{ \frac{\psi'(\a0)}{\psi(\a0)}\delta'(\tilde{r})
        - \frac{3}{\a0}
        \left[\frac{\psi'(\a0)}{\psi(\a0)}\delta'(\tilde{r})
        + \delta(\tilde{r}) 
        - \delta\left(\tilde{r} 
            + [\tilde\varrho + \tilde\psi(\a0)]/\a0
              \right)\right]
     \right\} + {\mathcal{O}}(\epsilon^2)\ . \nonumber
\end{eqnarray}

Since the above expression is first-order in $\epsilon$, we may
safely replace $\alpha_0$ by $\alpha(\tilde{r})$ because the
difference between the two is also first order, \equ{epsi}. 
Additionally, \equ{epsi} implies the identities
\begin{eqnarray}
\epsilon\delta'(\tilde{r}) &=& \alpha'(\tilde{r}) \ , \\
\epsilon\delta(\tilde{r}) - \epsilon\delta\left(\tilde{r}
+ [\tilde\varrho + \tilde\psi(\alpha_0)]/\alpha_0\right) &=&
\alpha(\tilde{r}) - \alpha\left(\tilde{r} 
+ [\tilde\varrho + \tilde\psi(\alpha_0)]/\alpha_0\right) \ ,
\end{eqnarray}
which allow us to rewrite \equ{pre-dalpha} in terms of 
$\alpha(\tilde r)$ and its first derivative $\alpha'(\tilde r)$ 
[with no other explicit functions of $\tilde r$ such as 
$\delta(\tilde r)$ or $\delta'(\tilde r)$]:
\begin{eqnarray}
\label{eq:a_final-first-order}
\Delta \alpha(\tilde{r}) &=&  
\frac{\alpha'(\tilde{r})}
{1 +\lambda^{-3} \varrho^{-3/\alpha} \psi(\alpha)^{1-3/\alpha}}\\
&&\times \left[\frac{\psi'(\alpha)}{\psi(\alpha)}
- \frac{3}{\alpha}
\left(\frac{\psi'(\alpha)}{\psi(\alpha)}
- \frac{\alpha\left(\tilde{r} 
+ [\tilde\varrho + \tilde\psi(\alpha)]/\alpha
\right) - \alpha}
{\alpha'(\tilde{r})}\right)\right] 
+ {\mathcal{O}}(\epsilon^2) \ . \nonumber 
\end{eqnarray}
Here, and in what follows, $\alpha$ stands for $\alpha(\tilde{r})$.
This notation hides the explicit dependence of
$\Delta\alpha(\tilde{r})$ on $\epsilon$, but for
the calculation to be self consistent $\epsilon$ has to be on the order of
\begin{equation}
\label{eq:order-epsilon} 
\epsilon \sim \frac{\alpha\left(\tilde{r}
+ [\tilde\varrho + \tilde\psi(\alpha)]/\alpha\right)
- \alpha}{\alpha} \ .
\end{equation}
We thus require the right-hand side of \equ{order-epsilon} to be much
smaller than unity.

In principle, the sign of $\Delta\alpha$ should allow an analysis of the
evolution of $\alpha$ into a fixed point. However, the third term in
\equ{a_final-first-order} is still problematic; it involves the function
$\alpha(\cdot)$ evaluated recursively at a point which depends on $\alpha$
itself. In order to make this term useful, one must make further
assumptions regarding the functional form of $\alpha(\cdot)$.

\subsection{The Linear Regime}

The most natural assumption for the functional form of
$\alpha(\tilde{r})$ is that it is linear in $\tilde{r}$.
As seen in Fig.~3 of DDH, this is clearly a good approximation for the
profile used in our merger N-body simulation, and it is also valid
for the family of profiles described in \equ{nfw}.
In other words, we
assume that the second and higher derivatives of $\alpha(\tilde{r})$ are
much smaller than the first derivative --- at least in the interval
$[\tilde{r}, \tilde{\ell_0} - \tilde{\lambda}]$. Under this assumption
we may approximate the above problematic term by
\begin{equation}
  \frac{\alpha\left(\tilde{r} 
     + [\tilde\varrho + \tilde\psi(\alpha)]/\alpha
     \right) 
    -\alpha(\tilde{r})} 
      {\alpha'(\tilde{r})} 
    \simeq \frac{1}{\alpha}
   [\tilde\varrho + \tilde\psi(\alpha)] \,,
\end{equation}
which brings \equ{a_final-first-order} to finally become
\begin{eqnarray}
  \Delta \alpha(\tilde{r}) &\simeq&  
    \frac{\alpha'(\tilde{r})}
         {1 + 
   \lambda^{-3} \varrho^{-3/\alpha} \psi(\alpha)^{1-3/\alpha} 
         } \nonumber\\    
    && \times \frac{1}{\alpha^2}
       \left(\alpha(\alpha - 3)\frac{\psi'(\alpha)}{\psi(\alpha)}
      + 3
[\tilde\varrho + \tilde\psi(\alpha)]
      \right) \,.
\label{eq:a_dalpha_l}
\end{eqnarray}

In this approximation, the sign of $\Delta \alpha(\tilde{r})$ is the
sign of the expression in big brackets, and the asymptotic fixed point
slope $\aas$ is obtained by solving an explicit equation of
$\alpha$:
\begin{equation}
      \alpha(\alpha - 3)\frac{\psi'(\alpha)}{\psi(\alpha)}
      + 3
    [\tilde\varrho + \tilde\psi(\alpha)]
      = 0 \,.
      \label{eq:a_aas}
\end{equation}
The factor in front of the big brackets of \equ{a_dalpha_l}
affects the rate at which $\alpha(\tilde r)$ converges to the asymptotic
slope. 
This concludes the derivation of \equ{dalpha_l} and \equ{aas} introduced
in \se{linear_summary} for the case of a given mass ratio $m/M$.

\section{A COSMOLOGICAL DISTRIBUTION OF SATELLITES}
\label{sec:dist}

The analysis so far referred to mergers with a fixed mass ratio $m$.
A real sequence of mergers consists of a cosmological mix of satellite masses.
The probability for a mass ratio $m$ in a merger
can be estimated using the Extended Press Schechter scheme 
(Lacey \& Cole 1993, EPS),
which has been tested and calibrated using N-body simulations 
(Lacey \& Cole 1994; Somerville \etal 1999, Fig.~2).
We have performed Monte Carlo realizations of merger trees based on EPS
(using a scheme developed by Yuval Birnboim, to be published elsewhere,
based on the scheme of Somerville \& Kolatt 1999),
and found that the probability distribution of $m$ is quite insensitive to the
actual power spectrum of initial fluctuations,
and it has no obvious time dependence. 
The fraction of number of mergers with mass ratio $m<1$ 
in the interval $(m,m+\dd m)$ can be approximated by 
\be
p(m)\, \dd m \propto m^{-3/2} \dd m \,.
\label{eq:p(m)}
\ee

The relevant distribution of mass ratio for the buildup of the inner profile
should also take into account the time it takes each satellite's orbit to decay
by dynamical friction from the halo virial radius $R$ to the central region.
The duration of the decay process can be estimated for circular orbits
in an isothermal halo (as in BT, eq.~7-26). We obtain
\be
\tfric = {1.17 \over m \ln(1/m)}{ R \over \vcirc}
\sim {0.18 \over m \ln(1/m)} \tini \,,
\label{eq:tfric}
\ee
where $\tini$ is the Hubble time when the merger starts with the satellite
at the halo virial radius,
and where an Einstein-deSitter cosmology is assumed at that time.
On one hand \equ{tfric} 
is an underestimate because we have ignored the weakening
effect of mass loss on the dynamical friction. 
If we adopt the mass-loss recipe $m(r) \propto M(r)$ (which can be deduced
from the resonance condition for isothermal halo and satellite) we obtain that
$\tfric$ becomes about twice as long as when mass loss is ignored.
On the other hand \equ{tfric} is
an overestimate in the case of eccentric orbits, where dynamical friction is
strong already at early stages of the merger. 
In our merger simulations we find that a satellite on a typical
orbit of eccentricity 1:6 decays to the center on a time scale  
shorter by a factor of $\sim 3$ compared to the decay time of a similar
satellite on a circular orbit.
Thus, the two effects roughly balance each other, and we can
adopt \equ{tfric} as a crude approximation.
Note that for $m\lsim 1$ the estimate in \equ{tfric} breaks down 
(e.g., the log term drives $\tfric$ to grow with $m$ for $m>e^{-1}$).
Since we wish to push the crude analysis to include such mergers despite the
breakdown of our approximations there, we simply keep $\tfric$ constant in 
this range.
 
The satellite is irrelevant for the cusp at $t_0$ if $\tfric (m) > t_0-\tini$,
which therefore defines a minimum relative satellite mass $\mmin$.
For larger $m$ we crudely multiply the mass function of \equ{p(m)}
by the approximate correction factor $\ffric(m)=[1-\tfric(m)/(t_0-\tini)]$.
For an order-of-magnitude estimate, we assume that a typical merger occurs at
$\tini = t_0/2$ (e.g., for an isothermal halo this is the typical time 
of collapse of the inner half mass of the halo).
This implies $\mmin \sim 0.067$ and $\ffric \sim (1-0.18/[m\ln(1/m)])$
for $\mmin\leq m \leq e^{-1}$.
A lower value for $\tfric$, 
or an earlier $\tini/t_0$,
would correspond to lowering the factor 0.18 in $\ffric$ and
therefore to a smaller $\mmin$. Since $\mmin\sim 0.067$ 
is already much smaller than unity, the effect of the actual value
of $\mmin$ on our final result is weak.

Given the corrected probability distribution $\ffric(m) p(m)$, 
we compute the average over $m$ of $\Delta \alpha (r,m)$ as
given in \equ{dalpha_l}. We obtain the asymptotic slope $\aas$ for the 
distribution of satellites by requiring that the average increment vanishes:
\be
\int_{\mmin}^1 \dd m \, \ffric(m)\, p(m)\, \Delta\alpha(r,m) = 0 \,,
\label{eq:aas_m}
\ee
namely,
\begin{eqnarray}
0 &=&
\int_{\mmin}^1 \dd m\, \ffric(m)\, p(m)\, 
\frac{1} {1 + m^{-1-\mnu(1-3/\alpha)} 
\psi(\alpha)^{1-3/\alpha} } \nonumber\\    
&& \times 
\left(\alpha(\alpha - 3)\frac{\psi'(\alpha)}{\psi(\alpha)}
+3 \ln[m^{-\mnu}\psi(\alpha)] \right)  \,.
\label{eq:a_aas_m}
\end{eqnarray}
Note that we have expressed $\varrho$ and $\lambda$ in terms of $m$ and
$\mnu$, using the scaling \equ{mnu}.
The factor $\alpha'(\tilde r)/\alpha^2$ has been taken outside the integral
and dropped from the equation because it is
independent of the stochastic variable $m$ of the current merger (though it is
affected by the mass ratios in previous mergers throughout the halo
history).

\begin{figure}[t]
\vskip 7cm
{\includegraphics{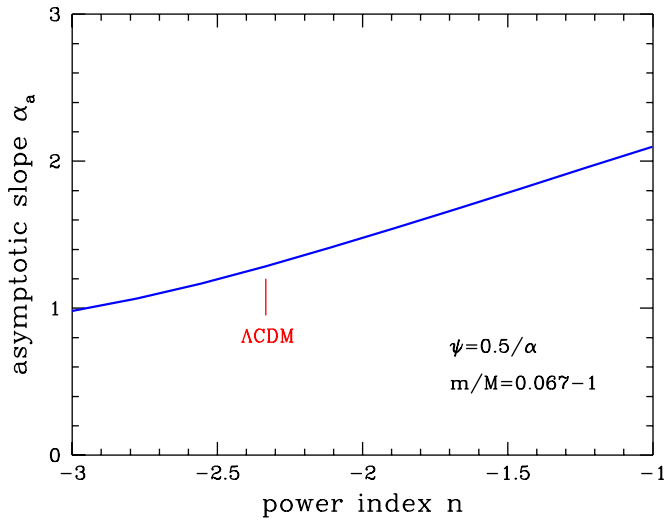}}
\caption{
The asymptotic slope due to a sequence of mergers with a cosmological
distribution of mass ratios, as obtained in the linear approximation
by solving \equ{a_aas_m}, shown
as a function of the fluctuation power index $n$.
}
\label{fig:am2}
\end{figure}

\Fig{am2} shows the value of the asymptotic profile obtained by solving
\equ{a_aas_m} for the above distribution of mass ratios 
and with $\psi(\alpha)=0.5/\alpha$
as a function of the power index $n$ of the fluctuation power spectrum
[$\nu=(3+n)/2$]. Two values are also marked as horizontal bars in \Fig{aas}.
For $\Lambda$CDM, with $n\simeq -2.33$, we obtain $\aas = 1.29$.
As seen in \Fig{aas}, the value of $\aas$ for a distribution of mass
ratios is similar to the values obtained from 
\equ{aas} for $m \simeq 0.32$ 
[which turns out to be close the average $\la m \ra \simeq 0.33$  
of the distribution $\ffric(m) p(m)$],
indicating that a typical merger in terms of its effect on the
cusp profile is with $m \simeq 1/3$.
This is true for any $n$.

The reason for the robustness of the dominance by 
$m \sim \la m \ra \simeq 1/3$
mergers and the insensitivity to the actual value chosen for $\mmin$ is the
relative flatness as a function of $m$ of the term multiplying the
expression in big parenthesis in the integrand, \equ{a_aas_m}.
The factor $\ffric(m)$ is roughly constant in most of the range,
while $p(m) \propto m^{-3/2}$.
The term expressed as a big fraction is roughly proportional to
$m^{1-\mnu(1-3/\alpha)}$ which is about $\propto m^{0.5}$ to $m$ 
in the relevant range of $\mnu$ and $\alpha$.
Together they introduce a moderately decreasing
dependence on $m$, which does not become too large even at small $m$,
as long as $\mmin$ is not significantly smaller than 0.01, say.

In \Fig{am2} we see a gradual variation of $\aas$ with $n$:
for the extreme case of $n=-3$ we get a minimum slope of $\aas = 0.98$ 
and for $n=-1.67$ we obtain $\aas=1.68$.
This range is comparable to the variation seen in simulations
of a single cosmology, e.g., $\Lambda$CDM.
We shall see below (\se{toy}) that the measured cusp slopes are actually 
expected to be somewhat less $n$-sensitive than $\aas$.
A systematic trend of such magnitude is hard to measure in cosmological
simulations of different power spectra.
There are hints for a similar trend already in the simulations of
Cole \& Lacey (1996, Fig.~9), and SCO report from their simulations
an average of $\ai =1.3\pm0.07$ for $n=-2$, then
$\ai =1.6\pm0.09$ for $n=-1$ and $\ai=1.8\pm0.09$ for $n=0$.

As a side complementary argument, we use a very simplistic toy model to
derive a crude upper limit for $\ai$. 
Following SCO,
we consider the accumulation of undigested satellites in the halo center.
Based on the assertion that the satellites' inner densities are significantly
higher than that of the host halo,
one assumes in this model that the satellites accumulating in the halo center 
keep their original profiles unchanged by tidal effects.
The mean halo profile that is built by a cosmological
distribution of satellites is then 
\be
\bar\rho(r) = \int_{\mmin} ^1 \dd m\, \ffric(m)\, p(m)\, \bar\sigma(m;r) \,
\label{eq:int_undigest}
\ee
where $\bar\sigma(m;r)$ is the profile $\bar\sigma(\ell)$ at $r=\ell$ of  
a satellite of mass ratio $m$, and where the $m$ dependence is given by the 
cosmological scaling for the given power spectrum. 
This integral is easy to evaluate if we assume that
$\ffric(m)={\rm const.}$, adopt $p(m) \propto m^{-3/2}$ as in \equ{p(m)}
based on the realizations of the EPS model, and crudely take 
$\bar\sigma(m;r)$ to be 
a step function of height $\sigc$ out to a radius $\ellc$, both scaled with
$m$ as in \equ{mnu}. At a given $r$, there is a contribution to $\bar\rho(r)$
from all satellites of $\ellc > r$, which translates by \equ{mnu} to
a lower bound for the integration at a minimum mass $m = r^{3/(1+\mnu)}$.
The result, with $\mnu = (3+n)/2$, is
\be
\ai = 3(4+n)/(5+n) .
\label{eq:undigest}
\ee
SCO argued for a flatter slope, $\ai = 3(3+n)/(5+n)$,
which could have resulted from \equ{int_undigest} if one had assumed
$\ffric(m)\,p(m) \propto m^{-1}$.
Based on the above discussion of dynamical friction we expect $\ffric(m)$
to be a moderately increasing function of $m$ for $m>\mmin$. 
This would flatten the slope predicted in \equ{undigest} and thus
make it an upper bound for the actual cusp slope. 
An effective behavior of 
$\ffric(m) \propto m^{1/3}$, with $p(m) \propto m^{-3/2}$,
would roughly recover the $n$ dependence found in the simulations of SCO.
This simplified toy model ignores the additional effects of tidal mass 
transfer [which we model in the current paper via $\psi(\alpha)$] 
and the associated effects of tidal distortion of the satellites' inner 
regions, which may go either way. 
Nevertheless, we notice that \equ{undigest} indeed overestimates our model 
prediction for $\aas$ for any $n$, and thus serves as an upper bound to $\ai$.

\section{TOY SIMULATIONS}
\label{sec:toy}

\begin{figure}[t]
\vskip 7cm
{\includegraphics{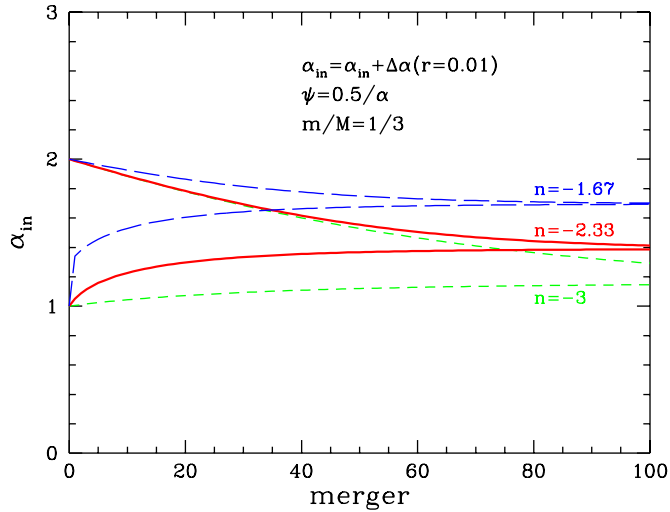}}
\caption{
Evolution of the cusp slope in a sequence of mergers with homologous
satellites of $m/M=0.33$. In this version of the toy simulation
the functional
form of \equ{nfw} is enforced with $\aout=3$, starting with either $\ain=1$
or 2. The change in $\ain$ is assumed to be given by $\Delta\alpha(r)$
at $r=0.01\rs$.
The cusp slope converges slowly to an asymptotic value, as predicted by the
linear analysis.
}
\label{fig:dal}
\end{figure}

In order to test the predictions of the linear analysis of \se{linear} for the 
asymptotic profile, and to study the evolution towards this asymptotic profile,
we perform toy simulations of the profile buildup by a sequence of mergers,
where we implement straightforwardly the mass-transfer recipe 
of \equ{psi} with $\psi(\alpha)=0.5/\alpha$ in the range 
$\alpha>1$.\footnote{Certain modifications to the recipe may be needed near 
and below $\alpha \sim 1$, where our adopted $\psi(\alpha)$ is not well 
established, and where deposit may occur without stripping. 
For example, if \equ{psi} yields a deposit radius $r<\ell$, 
one may wish to replace it by $r=\ell$, assuming that the surviving satellite 
supports itself in a finite configuration while its center sinks to the 
halo center.}
We start with a halo of a given mean density profile $\bar\rho(r)$ 
and the corresponding monotonically increasing $\alpha(r)$. 
We consider a merging satellite of mass ratio $m/M$ with a homologous profile
properly scaled using \equ{mnu}. 
To materialize the merger we solve \equ{psi} numerically for $\ell(r)$,
with the possible modification for deposit without stripping at $\alpha < 1$.
We then assume that the halo profile changes only due to the added stripped
satellite material, \equ{rho_final}, and compute the exact change in slope 
$\Delta\alpha(r)$ at any $r$ using \equ{dalpha}.
The following merger is performed with a satellite that is a scaled version of
the new halo profile, and so on.

As a first simple test, we enforce at all times a mean density profile obeying
the functional form of \equ{nfw}, with $\rs$ and $\aout$ fixed at 1 and 3
respectively. The inner slope $\ain$ starts at an arbitrary value in the range 
$1\leq\ain\leq 2$ and is allowed to change as a result of the merger.
We assume here that $\ain$ changes by $\Delta\alpha(r)$ as evaluated at some 
small radius $r$.  
\Fig{dal} shows the resultant evolution of $\ain$ for $m/M=0.33$ and 
$r=0.01\rs$, when starting alternatively from either $\ain=1$ or $\ain=2$, 
and for three different power spectra of fluctuations. 
We see that, indeed, there is a slow convergence to an asymptotic profile.
The obtained asymptotic slopes are close to the predictions of the
linear analysis, \Fig{aas}.
The slight deviations of less than 10\% are mostly due to the fact that 
we compute $\Delta\alpha$ at a finite small 
$r$ rather than at $r \rightarrow 0$.

We next perform a non-parametric toy simulation where we do
not enforce a specific functional form and do not constrain the outer slope
to remain steep. We only make sure that $\alpha(r)$
is monotonic and relatively smooth, as required for
the implementation of the deposit scheme.
At each step,
the mean density profile of the halo, $\bar\rho(r)$, is stored in an
array of shells spaced logarithmically (1000 shells, $\Delta \ln r = 0.05$),
and spline interpolated into any desired $r$.
The slope $\alpha(r)$ is computed via 4'th order spline interpolation.  
To make sure that $\alpha(r)$ remains monotonic and smooth, 
$\alpha(r)$ is first smoothed 
with a Gaussian of width $\Delta \ln r = 0.5$, and then interpolated 
inside intervals of $\Delta \ln r = 2$ (with an overlap
of $\pm 1$ with the neighboring intervals) using locally a functional
form for $\alpha(r)$ corresponding to \equ{nfw} with all the parameters free.
This is a general smoothing procedure that is not sensitive to the use
of this specific functional form in the local fit.
Our initial profile is again given by \equ{nfw},
with $\rc$ put at the center of the sampled $\ln r$ range.

The ultimate test of the linear approximation is shown in \Fig{aas},
where the results of the non-parametric 
toy simulations are presented by the symbols on top of the curves
describing the linear predictions.
We see that the actual sequence of mergers with a fixed mass ratio
$m/M$ converges to an asymptotic slope very similar
to that predicted by the linear approximation, for all values of $m/M$
in the range of interest. Significant deviations are obtained only
for very small $m/M$ values, below any realistic value of $\mmin$ as
implied by dynamical friction time-scales.

\begin{figure}[t]
\vskip 6.6cm
{\includegraphics{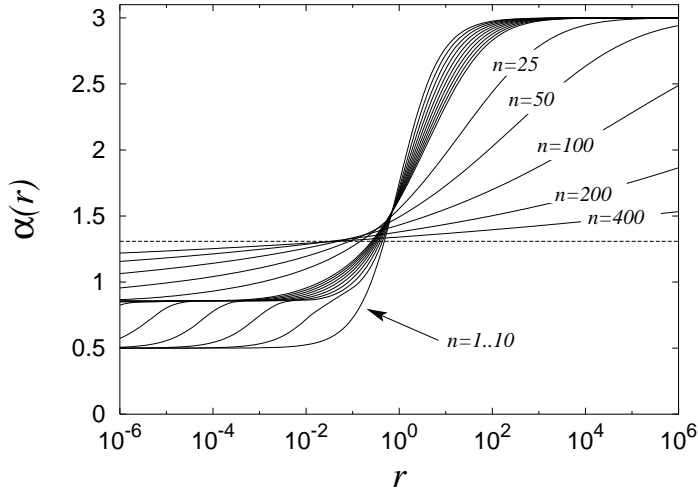}}
\caption{ 
Non-parametric and unconstrained toy simulation evolution of slope 
profile $\alpha(r)$
due to a sequence of mergers ($n=1,400$) with mass ratio $m/M=0.3$.
The initial profile is \equ{nfw} with $\ai=0.5$ and $\ao=3$.
A power-law region develops below the radius where $\Delta\alpha=0$,
with a slope that grows slowly from near unity to the asymptotic value.
For a long time, the profile maintains a shape of an inner cusp and a
steeper outer halo, as seen in cosmological simulations.
}
\label{fig:alpha_prof}
\end{figure}

\Fig{alpha_prof} shows how $\alpha(r)$ actually evolves in the non-parametric
toy simulations where no constraints are applied on the outer slope.
When $\ain$ is below unity there is a rapid evolution towards $\alpha \sim 1$
in the range $r < \rc$. This is partly due to the factor $1/\alpha^2$ driving
$\Delta\alpha$ in \equ{dalpha} and partly due to the processes explained in DDH
as reflected in the shape adopted for $\psi(\alpha)$ at $\alpha<1$.
A power-law region with $\alpha\sim 1$ develops below the radius where 
$\Delta\alpha=0$, and it quickly extends to smaller radii.
The value of $\alpha$ in this inner power-law region grows 
slowly from near unity towards the asymptotic value $\aas$. 
In parallel, since no constraints are applied on the outer slope,
it can decrease slowly from $\aout=3$ towards $\aas$.
The shape of the profile during the long interim period resembles 
a cusp-halo structure qualitatively similar to the NFW-like profiles of halos
seen in cosmological simulations.
It takes 10 to 20 mergers (of $m/M=0.33$), during which the halo mass grows by
about two orders of magnitude, for the shape of the profile to strongly deviate
from this characteristic shape. 
In reality, we may expect the outer profile to keep a steep slope in response
to other physical processes not modeled here (see \se{intro}), 
and therefore the asymptotic
profile to resemble a shape qualitatively similar to \equ{nfw} 
with $\ain=\aas$.

\section{DISCUSSION AND CONCLUSIONS}
\label{sec:conc}

We have addressed the origin of a cusp in dark-matter halos 
within the hierarchical clustering scenario, 
based on tidal effects during the halo buildup by merging 
of relatively compact 
satellite halos.  In Dekel, Devor \& Hetzroni (2003) we highlight the
rapid steepening of a flatter core of slope $\alpha \leq 1$ into a cusp of 
$\alpha >1$ as a result of vanishing tidal mass transfer in such a core. 
Merger N-body simulations indicate here that the cusp is quite stable under 
a sequence of such mergers. 
we have developed a simple prescription
for tidal mass transfer in mergers, argued using N-body simulations
that it provides a useful approximation, and showed that it 
naturally leads to a stable cusp with an asymptotic profile of slope $\aas$
slightly larger than unity.  Our toy model 
is one way to explain 
the cusps seen in cosmological N-body simulations
of the hierarchical clustering scenario, 
and it provides a tool for addressing other processes which may
explain the observed flat core in some galaxies.

The mass-transfer recipe relates each radius $\ell$ in the initial satellite
with a deposit radius $r$ in the halo via the ratio of mean densities
$\bar\rho(r)/\bar\sigma(\ell)=\psi[\alpha(r)]$. We argued based on qualitative
considerations and found in merger N-body simulations that $\psi(\alpha)$ 
has a characteristic shape which is not too sensitive to the details of 
the merger orbit; $\psi(\alpha)$ is a decreasing function of $\alpha$
and it obtains values significantly smaller than unity.
A crude fit seems to be provided by $\psi(\alpha)=0.5/\alpha$ in the range
$1\leq\alpha\leq 3$.
This makes the mass transfer more efficient than what might be inferred
from the naive resonance condition, $\psi=1$, and gradually more efficient
where the halo profile is steeper.  
Beyond our immediate purpose of studying the formation of a cusp,
this mass-transfer recipe provides a useful tool for general studies 
using semi-analytic models of galaxy formation.

Admittedly, we have only tested in this paper the robustness of 
$\psi(\alpha)$ to the merger orbit type.
The choice of typical properties for the simulated halo and satellite,
which relate to each other in general agreement with the scaling relation 
expected in the $\Lambda$CDM cosmology, makes the obtained
$\psi(\alpha)$ a sensible crude guess. Nevertheless, the robustness of
$\psi(\alpha)$ to variations about the simulated properties of
halo and satellite must be tested using a more comprehensive set of 
simulations. We have adopted $\psi(\alpha)=0.5/\alpha$ as an illustrative
example for obtaining actual numerical values for $\aas$, but the convergence 
analysis, which is the heart of this paper, is much more general.

After understanding why a decreasing $\psi(\alpha)$ should lead to a stable
asymptotic profile, we have performed a linear perturbation analysis 
of this process, and confirmed its validity by toy simulations of
tandem mergers based on the adopted mass-transfer recipe.
The asymptotic slope $\aas$ is typically somewhat larger than unity.
The system develops a cusp of $\alpha \sim 1$ which slowly grows towards
$\aas$ while the outer halo maintains a steeper profile for a long time.
For a typical cosmological distribution of merging satellites
in the $\Lambda$CDM cosmology ($n=-2.33$) the asymptotic slope is 
$\aas \simeq 1.3$.  This is close to the typical cusp slope found in
cosmological N-body simulations (e.g. Power \etal 2002 and other references in 
\se{intro}).
The typical mergers driving this process are
with a relatively high mass ratio of $m/M \sim 1/3$. 
This means that one should not take our
quantitative results too literally because our toy model assumes $m\ll M$.

The asymptotic profile obtained in our model has a relatively weak dependence
on the power spectrum of fluctuations.  In particular,  
the effect of tidal compression (\se{tide}) and the $1/\alpha^2$ factor in
$\Delta\alpha$ [\equ{dalpha}] drive the profile to $\alpha \gsim 1$
independently of the fluctuation power spectrum. Also,   
the cosmological dependence of the ultimate asymptotic slope is weakened
by the fact that the profile is determined by mergers of relatively
large mass ratio.
For values of power-spectrum index in the range $-3 < n < -1$ the asymptotic
slope is predicted to vary gradually in the range $1 < \aas < 2$.
This can be compared to the only slightly weaker trend indicated in the 
cosmological N-body simulations of SCO. 

Syer \& White (1998) also addressed the profile resulting from mergers
and obtained results somewhat different from ours. 
They adopted the condition of tidal stripping at resonance,
$\psi=1$, ignoring the $\alpha$ dependence of $\psi$.
When we substitute the recipe $\psi=1$ in \equ{dalpha}, we find that
$\Delta\alpha$ is always positive, for any merger and at any $r$.
Indeed, when trying to repeat the Syer \& White toy simulations
with higher resolution and following more mergers we actually find that
the profile does not really converge to a stable cusp but rather
continue to steepen slowly towards $\ai=3$. 
Only when using the revised mass-transfer prescription
where $\psi$ is decreasing with $\alpha$
do we obtain convergence to a flatter asymptotic profile.

Syer \& White (1998), Nusser \& Sheth (1999) and Subramanian, Cen \& Ostriker
(2000) describe different toy models which 
predict $\ai=(9+3n)/(5+n)$,\footnote{Nusser \& Sheth also predict an upper 
bound of $\ai=(9+3n)/(4+n)$.} recovering the hierarchical-clustering
toy model by Peebles (1980, eq.~26.8). 
While this scaling relation may be in reasonable agreement 
with the simulations (and with our model predictions) for $n \geq -1$,
it admits values below $\ai=1$ for $n<-2$, in conflict with the
findings of the simulations.  Our revision of the toy model of SCO
yields $\ai < 3(4+n)/(5+n)$, to be flattened further by the $m$ dependence 
of dynamical friction. The basic model analyzed in the current paper does 
not admit cores significantly flatter than $\ai \sim 1$ (that is
by non-dissipative 
processes only) and is thus in better agreement with the simulation results. 

The gravitational processes leading to a cusp are modeled in our analysis
as tidal effects during the buildup of the halo by a sequence of {\it mergers}.
This picture is likely to be valid in the CDM hierarchical clustering scenario,
where numerous sub-galactic halos exist on all scales and are continuously 
merging (e.g., Klypin \etal 1999b; Moore \etal 1999a; Springel \etal 2001). 
This merger picture
is confirmed by a careful inspection of high resolution CDM simulations,
where a special effort is made to identify the merging clumps which otherwise
could have been easily missed (e.g., Wechsler, Dekel \etal~, in preparation).
It is therefore clear that cores, independently of how they form,
cannot survive in a pure CDM scenario without some modification --- they
efficiently turn into cusps as a result of the mergers with the relatively
compact building blocks.

Nevertheless, a cusp, though somewhat flatter, is reported to be seen also in
simulations where the 
initial fluctuations had less power on small scales, thus suppressing the
number of sub-galactic satellites and the associated merger rate
(Moore \etal 1999b; Avila-Reese \etal 2001; Bullock, Kravtsov \& Colin 2002).
Further indications for the generality of cusp formation comes
from simulations by Huss, Jain \& Steinmetz (1999), who find that cusps
also form as a result of collapse from 
spherical initial density perturbations with 
suppressed 
random velocity perturbations, 
as well as from the
simulations of Alvarez, Shapiro \& Martel (2002), who find that
cusps arise from the gravitational instability and fragmentation
of cosmological pancakes.
One way to explain this is by noticing that
the asymptotic cusp formed in the CDM scenario is driven by mergers
with relatively massive satellites (of typical mass ratio 1:3, 
see Dekel \etal 2002), and realizing that such mergers do happen even
when small-scale power is suppressed and when pancakes fragment. 
It is therefore possible that the cusp is actually driven by mergers 
to a certain extent also in 
some of 
these cases. 
Our toy model based on mergers would not be directly applicable in cases
where a cusp arises while the halo forms by smooth accretion.
The gravitational processes involved in such cases somehow mimic a behavior 
similar to the merger case.  We note, for example, that the tidal compression 
discussed in DDH is expected to amplify density perturbations and possibly 
make them behave in certain ways like merging clumps.

However, one should accept in general that 
the halo buildup is a complex gravitational process, whose different aspects
can probably be modeled in more than one way,  
e.g., as a violent relaxation process driven by fluctuations in comparison
with a sequence of mergers and substructure accretion.\footnote{We 
encounter an analogous duality, for example, when one manages to explain
the origin of galactic spin alternatively via tidal torque theory applied 
to shells and by summing up the orbital angular momenta in a cosmological
sequence of mergers (see Maller, Dekel \& Somerville 2002).} 
There is also the intriguing possibility that general statistical
considerations in phase space may provide some clue for the origin of
a universal halo profile (e.g., Taylor \& Navarro 2001). 
Unfortunately, we currently know of no viable alternative to the tidal effects
in mergers as a simple model for
the origin of $\alpha \gsim 1$ cusps in the cosmological simulations.
The idealized merger picture provides one possible toy model, 
approximately valid at least in hierarchical clustering scenarios,
within which we understand the origin of the cusp in simple terms.
However, the complexity of the problem suggests that this is not the final 
word on the issue.

Our result implies a {\it necessary}
condition for the survival of cores in halos independent of their origin
--- that satellites should be prevented from adding mass to the halo cores.
This could be avoided in a CDM scenario if feedback processes manage to puff up
the satellites and make them disrupt before they merge with the halo cores
(as proposed in DDH).
We have not explicitly addressed in this paper the {\it sufficient} 
conditions for 
the formation of cores, but one can imagine that a sequence of mergers with
low-density satellites, where the mass is predominantly deposited 
outside the inner halo region, would indeed flatten the inner profiles
(work in progress).
Other processes may also contribute to the development of halo cores. 
The proposed scenarios include, for example,
the disruption of cusps in merging satellites by massive black holes 
(Merritt \& Cruz 2001),
the heating by gas clouds spiraling in due to dynamical friction
(El-Zant, Shlosman \& Hoffman 2002),
the angular-momentum transfer from a big temporary rotating bar 
(Weinberg \& Katz 2002),
and the delicate resonant reaction of halo-core orbits to the tidal
perturbation by the satellite, which could be a strong effect if the
dark-matter distribution is much smoother than the current state-of-the-art
N-body simulations (Katz \& Weinberg 2002).
We stress again that, no matter what the origin of the core might be,
our analysis implies that such cores could survive only if  
they are not perturbed by significant mass transfer from merging
compact satellites, which implies that small CDM halos must be puffed up
before they merge into bigger halos. We also stress that none of these
scenarios seem to be capable of explaining the formation of cores in 
halos on the scales of clusters of galaxies within the CDM scenario.

Supernova feedback effects are probably not strong enough for turning
cusps into cores in halos with rotation velocities higher than $\sim 100\kms$
(Geyer \& Burkert 2001; Gnedin \& Zhao 2002). However, they
are possibly sufficient for the necessary indirect puffing-up of the merging
satellites.
Gnedin \& Zhao (2002) estimate that direct feedback effects may
reduce the central satellite densities by a factor of 2 to 6. Based on our
simulations with puffed-up satellites, this may be enough by
itself to avoid the steepening from a core to a cusp in the framework of the
merging scenario.
However, if the (so far inconclusive) observational clues for cores in
clusters of galaxies are confirmed,
supernova 
feedback is unlikely to provide a viable explanation for their origin.
In this case one may search for higher efficiency in supernova feedback
either due to microscopic effects such as porosity
in a muliphase ISM or due to hypernova from very massive stars (Silk 2002).
Alternatively, one may appeal to stronger feedback mechanisms, perhaps
associated with radio jets from AGNs. This process may be indicated
by an observed correlation between AGN activity and bright galaxies
in SDSS (Kauffmann \etal 2003, in preparation),
and it may be needed independently in order to explain the missing baryons
in big galaxies (Klypin, Zhao \& Somerville 2002) and in clusters.
Otherwise, such large cores may present a real challenge for the standard
hierarchical clustering scenario

The cusp/core problem is only one of the difficulties facing galaxy formation
theory within the CDM cosmology. It turns out that other main
problems can also be modeled by tidal effects in mergers, and may also be
resolved by the inevitable feedback processes. For example,
Maller \& Dekel (2002) addressed the
angular-momentum problem, where simulations including gas
produce disks smaller than the galactic disks observed
(Navarro \& Steinmetz 2000 and references therein; Governato \etal 2002),
and with a different internal distribution of angular momentum
(Bullock \etal 2001b; van den Bosch, Burkert \& Swaters 2001).
A toy model has been constructed for the angular-momentum buildup by mergers
based on tidal stripping and dynamical friction, which helps us understand
the origin of the spin problem as a result of over-cooling in satellites.
A simple model of feedback has then been incorporated,  
motivated by Dekel \& Silk 
(1986). This model can remedy the discrepancies, and in particular
explain simultaneously the low baryon fraction and angular-momentum
profiles in dwarf disk galaxies.

Various feedback effects may also provide the cure to the missing dwarf 
problem,
where the predicted large number of dwarf halos in CDM can possibly match
the observed number of dwarf galaxies only if the mass-to-light ratio
in these objects is very high (Klypin \etal 1999b; Moore \etal 1999a;
Springel \etal 2001; Kochanek 2001).
Bullock, Kravtsov \& Weinberg (2000), Somerville (2002) and 
Tully \etal (2002) 
appeal to radiative feedback effects which prevent the formation of
small dwarfs after cosmological reionization at $z\sim 7$,
Scannapieco, Ferrara \& Broadhurst (2000) and Scannapieco \& Broadhurst (2001)
address the destructive effect of outflows from one galaxy 
on neighboring protogalaxies via ram pressure,
and Dekel \& Woo (2002) study the role of supernova feedback in determining
the relevant global properties of dwarfs and larger low-surface-brightness
galaxies.
We note that
while the requirements from feedback in explaining the dwarf-galaxy 
properties and the angular-momentum problem are not too demanding, the
solution to the core problem requires that the dark-matter distribution
be affected by feedback, which is a non-trivial requirement.

Nevertheless, 
the successes of such toy models in matching several independent observations
indicate that they indeed capture the relevant basic elements of the complex
processes involved, and in particular that feedback effects may indeed
provide the cure to some or all the main problems of galaxy formation theory
within the $\Lambda$CDM cosmology that does so well on larger scales.
The alternative solution involving Warm Dark Matter (e.g., Hogan \& Dalcanton
2000; Avila-Reese \etal 2001; Bode, Ostriker \& Turok 2001) 
seems to still suffer 
from the 
cusp 
problem, it may still fail
to reproduce the angular-momentum profile in galaxies
(Bullock, Kravtsov \& Colin 2002), and it may be an 
overkill where the formation of dwarf galaxies is totally suppressed once the
inevitable feedback effects are included (Bullock 2001).
The speculative alternatives involving self-interacting dark matter
are even more problematic 
(Spergel \& Steinhardt 2000; Dave \etal 2001; Hennawi \& Ostriker 2002).

\section*{Acknowledgments}
We acknowledge stimulating discussions with Oded Ben David,
George Blumenthal, Andi Burkert, Doug Lin, Ari Maller and Gary Mamon.
This research has been supported 
by the US-Israel Bi-National Science Foundation grant 98-00217,
the German-Israel Science Foundation grant I-629-62.14/1999,
and NASA ATP grant NAG5-8218.

\def\refe{\reference}

\end{document}